\documentclass[journal]{IEEEtran}
\usepackage{graphicx}

\usepackage{caption}
\usepackage[caption=false,font=footnotesize]{subfig} % IEEE caption 스타일 유지

\usepackage{amsthm}
\usepackage{cite}
\usepackage[utf8]{inputenc}

\ifCLASSINFOpdf
\else
\fi
\usepackage{booktabs}
\usepackage{tabularx}
\usepackage{array}

\usepackage{amsmath}
\usepackage{xcolor}

\usepackage{algorithmic}
\usepackage{algorithm}
\usepackage{bm}
\usepackage{graphicx}
\usepackage{booktabs}
\usepackage{array}

\usepackage{amssymb}

\begin{document}
%
% paper title
% Titles are generally capitalized except for words such as a, an, and, as,
% at, but, by, for, in, nor, of, on, or, the, to and up, which are usually
% not capitalized unless they are the first or last word of the title.
% Linebreaks \\ can be used within to get better formatting as desired.
% Do not put math or special symbols in the title.
%\title{Robust Beamforming Design Considering Imperfect CSI for Time-Division Cell-Free Near-Field ISAC}
% Time-Division Cell-Free Near-Field ISAC with Robust Beamforming and Time Allocation

%\title{Robust Beamforming Design Considering Sensing Error for Time-Division Cell-Free Near-Field ISAC}
\title{Joint RF--VLP Position and Velocity Tracking: Bayesian Estimation and Performance Analysis}
%
%
% author names and IEEE memberships
% note positions of commas and nonbreaking spaces ( ~ ) LaTeX will not break
% a structure at a ~ so this keeps an author's name from being broken across
% two lines.
% use \thanks{} to gain access to the first footnote area
% a separate \thanks must be used for each paragraph as LaTeX2e's \thanks
% was not built to handle multiple paragraphs
%

\author{Chaedam~Son and Si-Hyeon Lee,~\IEEEmembership{Senior Member,~IEEE}
        % <-this % stops a space
\thanks{The authors are with the School of Electrical Engineering, Korea Advanced Institute of Science and Technology (KAIST), Daejeon 34141, South Korea (e-mail: \{scd5929,sihyeon\}@kaist.ac.kr; corresponding author: Si-Hyeon Lee).}% <-this % stops a space% <-this % stops a space
%\thanks{Manuscript received April 19, 2005; revised August 26, 2015.}
}

% note the % following the last \IEEEmembership and also \thanks - 
% these prevent an unwanted space from occurring between the last author name
% and the end of the author line. i.e., if you had this:
% 
% \author{\dots.lastname \thanks{\dots} \thanks{\dots} }
%                     ^------------^------------^----Do not want these spaces!
%
% a space would be appended to the last name and could cause every name on that
% line to be shifted left slightly. This is one of those "LaTeX things". For
% instance, "\textbf{A} \textbf{B}" will typeset as "A B" not "AB". To get
% "AB" then you have to do: "\textbf{A}\textbf{B}"
% \thanks is no different in this regard, so shield the last } of each \thanks
% that ends a line with a % and do not let a space in before the next \thanks.
% Spaces after \IEEEmembership other than the last one are OK (and needed) as
% you are supposed to have spaces between the names. For what it is worth,
% this is a minor point as most people would not even notice if the said evil
% space somehow managed to creep in.

\allowdisplaybreaks[4]

% The paper headers
\setlength{\arraycolsep}{5pt}    % 가로 간격 (기본 ~5pt 정도)
\renewcommand{\arraystretch}{0.9}% 세로 간격 (기본 1.0)
% The only time the second header will appear is for the odd numbered pages
% after the title page when using the twoside option.
% 
% *** Note that you probably will NOT want to include the author's ***
% *** name in the headers of peer review papers.                   ***
% You can use \ifCLASSOPTIONpeerreview for conditional compilation here if
% you desire.

% If you want to put a publisher's ID mark on the page you can do it like
% this:
%\IEEEpubid{0000--0000/00\$00.00~\copyright~2015 IEEE}
% Remember, if you use this you must call \IEEEpubidadjcol in the second
% column for its text to clear the IEEEpubid mark.

% use for special paper notices
%\IEEEspecialpapernotice{(Invited Paper)}
\newtheorem{remark}{Remark}
\newtheorem{theorem}{Theorem}
\newtheorem{lemma}{Lemma}
\newtheorem{Corollary}{Corollary}
\newtheorem{proposition}{Proposition}
% make the title area
\maketitle

% As a general rule, do not put math, special symbols or citations
% in the abstract or keywords.
\begin{abstract}
Visible light positioning (VLP) provides accurate indoor positioning through optical received signal strength (RSS) measurements,
while radio-frequency (RF)  delay and Doppler measurements offer
complementary range and velocity information. 
Motivated by this complementarity, we investigate
joint RF--VLP position and velocity tracking,
addressing both recursive state estimation and its
fundamental performance characteristics.
We formulate a Bayesian maximum a posteriori (MAP)
estimation problem that combines optical RSS,
RF delay and Doppler measurements, and motion
prediction, and develop a Gauss--Newton-based
algorithm for recursive tracking.
To characterize the estimation limits and the
benefits of RF--VLP integration, we derive the
Bayesian information matrix and the corresponding
Bayesian Cram\'er--Rao lower bound (BCRLB).
We further introduce a state-conditioned information
matrix (SCIM), whose inverse coincides with the
MAP estimator's mean-squared error matrix under
a first-order observation model, enabling tractable
analysis of RF-induced information gains.
The analysis quantifies the distinct contributions
of RF delay and Doppler measurements and reveals
different asymptotic estimation-error behaviors
with respect to RF and VLP signal-to-noise ratios.
Numerical results demonstrate improved position
and velocity tracking over the considered baselines,
with substantial velocity gains from RF sensing.
The results also validate the theoretical
asymptotic predictions and show that the inverse
of the SCIM closely captures the estimator's
error trends. 
\end{abstract}

% \begin{abstract}
% Visible light positioning (VLP) can provide highly accurate indoor
% position information but lacks direct Doppler measurements for velocity
% estimation, whereas radio-frequency (RF) sensing can directly provide
% Doppler information. Motivated by these complementary characteristics,
% we propose an RF--VLP framework that augments VLP measurements with RF
% delay and Doppler information for joint User position and velocity
% tracking. A state-transition model is incorporated to
% capture user motion, and the resulting maximum a posteriori estimation
% problem is solved using a Gauss--Newton-based algorithm. We further
% derive the Bayesian information matrix and the corresponding Bayesian
% Cramér--Rao lower bound (BCRLB). For tractable performance analysis, we
% introduce a state-conditioned information matrix (SCIM) and characterize
% the RF-induced information gains and the asymptotic behavior of the SCIM
% and its inverse with respect to RF and VLP signal-to-noise ratios (SNRs). Numerical results show that the proposed estimator outperforms
% the considered estimation baselines and that RF--VLP integration
% yields lower position and velocity error measures than the
% VLP-only configurations. The measured root mean square errors remain above the
% BCRLB-based error bounds, while the inverse-SCIM-based measures
% closely capture their trends.
% \end{abstract}


% For peer review papers, you can put extra information on the cover
% page as needed:
% \ifCLASSOPTIONpeerreview
% \begin{center} \bfseries EDICS Category: 3-BBND \end{center}
% \fi
%
% For peerreview papers, this IEEEtran command inserts a page break and
% creates the second title. It will be ignored for other modes.
\IEEEpeerreviewmaketitle

\section{Introduction}
\IEEEPARstart{W}{ith} the growing demand for location-based services,
accurate positioning has become increasingly important for applications
such as autonomous robots, user localization, and smart factories
\cite{ISAC_survey2,ISAC_survey3}.
However, conventional satellite-based positioning technologies,
such as the Global Positioning System (GPS) and real-time kinematic (RTK)
positioning, are severely limited in indoor environments due to poor
satellite visibility and complex propagation conditions
\cite{indoor_survey}.
To address these limitations, radio frequency (RF)-based indoor
positioning has been extensively investigated
\cite{RF_indoor1,RF_indoor4, RF_indoor3,single_anchor_positioning}.
By exploiting RF propagation characteristics,
\cite{RF_indoor1} combined time-of-arrival and angle-of-arrival
measurements in an ultra-wideband system for accurate localization,
while \cite{RF_indoor4} utilized Wi-Fi phase measurements
to infer position and motion information with decimeter-level accuracy.
At the system level, \cite{RF_indoor3} developed a hierarchical localization architecture using relay stations and a few base stations.
From a theoretical perspective,
\cite{single_anchor_positioning} derived performance bounds
for joint position and velocity estimation using a single RF anchor
and characterized their asymptotic behavior.
 
These studies demonstrate the potential of RF-based positioning
as an alternative to GPS and RTK in indoor environments.
Meanwhile, visible light positioning (VLP) has attracted
considerable attention as another promising indoor positioning
technology, benefiting from predominantly line-of-sight (LoS)
propagation and multiple spatially distributed LEDs.
In particular, received signal strength (RSS)-based VLP exploits the position-dependent
received optical power to enable accurate localization
with relatively simple receiver implementation
\cite{VLP_review1,VLP_review2}.
Building on these advantages, various RSS-based VLP
methods have been developed for different indoor
localization scenarios.
In \cite{VLP-CS}, a compressed-sensing-based framework
was proposed for simultaneous localization of multiple
robots, exploiting both inter-target and inter-anchor
cooperation and incorporating a generative adversarial
network (GAN) to enhance robustness.
Another collaborative VLP approach was investigated
in \cite{VLP_Collaborative}, where multiple targets
utilize LED signals and inter-target infrared
measurements, together with theoretical performance
analysis.
In \cite{VLP_Distributed_collaborative},
a distributed message-passing-based approach
was developed for collaborative localization
of mobile targets. 
Meanwhile, \cite{vlp_shadowing} investigated
passive target localization based on RSS changes
induced by shadowing, jointly considering
sensor position calibration and deriving
the corresponding performance bounds.

The high positioning accuracy demonstrated by VLP
makes it attractive for indoor tracking.
However, optical RSS measurements depend on the
receiver position and do not directly provide
velocity information.
Consequently, velocity estimation in RSS-based VLP
relies on observations collected
over time 
\cite{VLP_Distributed_collaborative,vlp_shadowing}.
In contrast to optical RSS, RF Doppler measurements
provide direct information about radial velocity,
while RF delay measurements provide range information.
These complementary measurement capabilities
motivate the integration of RF sensing with VLP
for joint position and velocity tracking.
Despite this potential, existing RF--VLP integration
studies \cite{RF+VLP1,RF+VLP2,RF+VLP3}
have primarily employed RSS-based RF measurements
to improve indoor positioning under limited
optical coverage or anchor availability.
Specifically, \cite{RF+VLP1} incorporated
Wi-Fi RSS fingerprinting into a crowdsourced
multisignal localization framework,
while \cite{RF+VLP2} combined RF-based
fingerprinting with VLP to facilitate position initialization
under unfavorable LED geometries. 
More recently, \cite{RF+VLP3} utilized
inter-receiver RF RSS measurements to facilitate
localization with sparse LED anchors.
Although these studies demonstrate the benefits
of RF--VLP integration for positioning,
they did not exploit RF delay and Doppler
measurements for joint position and velocity
tracking.
Such joint tracking also calls for a theoretical
performance analysis to establish the fundamental
benefits of RF--VLP integration.
In particular, this analysis can reveal
how RF measurements contribute additional
information  
and characterize the resulting limits on position
and velocity estimation accuracy.

Motivated by this need, we investigate
joint RF--VLP position and velocity tracking
using optical RSS and RF delay and Doppler
measurements, and develop a theoretical framework
to quantify the resulting estimation performance
gains.
The main contributions of this work are
summarized as follows.

\begin{itemize}

\item We develop a Bayesian framework for joint position
and velocity tracking that integrates RF delay and Doppler
measurements with VLP RSS observations.
We formulate a recursive maximum a posteriori (MAP)
estimation problem based on the joint measurement
and motion models and develop a Gauss--Newton-based
algorithm to solve the resulting nonlinear problem.

\item We establish fundamental performance limits and characterize the information gains of joint
RF--VLP position and velocity tracking.
We derive the Bayesian information matrix (BIM)
and the corresponding Bayesian Cramér--Rao lower
bound (BCRLB), and introduce a state-conditioned
information matrix (SCIM) for tractable analysis.
Under a linearized measurement model, we establish a relationship between
the inverse of the SCIM and the MAP estimator's
mean-squared error (MSE) matrix.
Using the SCIM, we quantify the information gains
provided by RF measurements and characterize
the asymptotic behavior of the SCIM and its inverse
with respect to RF and VLP signal-to-noise ratios (SNRs).

\item We numerically validate the proposed tracking
framework and the theoretical performance analysis.
The proposed estimator achieves lower position
and velocity root mean square errors (RMSEs) than the considered baseline
algorithms, while the inverse-SCIM-based error
measures closely follow the RMSE trends,
particularly under model-consistent motion.
Moreover, RF--VLP integration improves both position
and velocity estimation over VLP-only configurations,
with velocity gains persisting even as the number
of LEDs increases, highlighting the distinct
benefits of RF Doppler measurements.
The numerical results also corroborate the
theoretically predicted asymptotic behaviors
with respect to RF and VLP SNRs.
\end{itemize}

\textit{Notation}: Scalars, vectors, and matrices are denoted by regular,
bold lowercase, and bold uppercase letters, respectively, e.g.,
\(x\), \(\mathbf{x}\), and \(\mathbf{X}\).
The identity, zero vector, and zero matrix are denoted by
\(\mathbf I_N\), \(\mathbf 0_N\), and \(\mathbf 0_{M\times N}\),
respectively. The operators \(\operatorname{diag}(\cdot)\) and
\(\operatorname{vec}(\cdot)\) denote diagonal-matrix construction and vectorization,
while \((\cdot)^T\) and \((\cdot)^H\) denote transpose and Hermitian
transpose. The symbols \(|\cdot|\) and \(\|\cdot\|\) denote absolute
value and the Euclidean/Frobenius norm, respectively.
The Gaussian and circularly symmetric complex Gaussian distributions
are denoted by \(\mathcal N(\pmb{\mu},\pmb{\Sigma})\) and
\(\mathcal{CN}(\pmb{\mu},\pmb{\Sigma})\), with their probability
density functions written as
\(\mathcal N(\mathbf{x};\pmb{\mu},\pmb{\Sigma})\) and
\(\mathcal{CN}(\mathbf{x};\pmb{\mu},\pmb{\Sigma})\), respectively.
Finally, \(\mathbb E[\cdot]\) denotes expectation.
%\begin{figure}[t]
%    \centering
%    \includegraphics[width=6cm]{vlp_system_model_figure.png}
%    \caption{Joint RF–VLP indoor tracking system.}
%    \label{main}
%\end{figure}

% Recommended notation tables for:
% Joint RF--VLP Position and Velocity Tracking: Bayesian Information Analysis and Signal-Level Fusion
% Required packages:
% \usepackage{booktabs,tabularx,array}
% Optional for tighter spacing: \usepackage{makecell}

% IMPORTANT CLEANUP BEFORE INSERTION
% 1) Use \eta_t^{(\ell)} (not \beta) for the GN step size.
% 2) Use \ell only as the GN iteration counter; do not use r as a second counter.
% 3) In the state-conditioned analysis and Appendices C--F, use \mathbf J_t^{\mathrm c}
%    and \mathbf P_t^{\mathrm c}, not \mathbf J_t^{\mathrm B} or ``CRLB''.
% 4) Use \widetilde\Gamma_t\triangleq 8\pi^2\Gamma_t^{\mathrm{RF}} consistently
%    wherever the factor 8\pi^2 has been absorbed.
% 5) The proof references should be: Theorem 2 -> Appendix C; Theorem 3 -> Appendix D;
%    Theorem 4 -> Appendix E; Theorem 5 -> Appendix F.
\section{System Model} \label{sec:model}
We consider an indoor RF--VLP tracking system, where a user equipment (UE) estimates its own position
and velocity using reference signals from \(M\geq3\) LEDs and one RF transmitter. This configuration represents a practical indoor setting in which multiple spatially distributed LED luminaires are available for illumination, while a single RF transmitter is used to complement the VLP measurements. The UE is equipped with a co-located photodetector (PD) and RF antenna
for visible light (VL) and RF signal reception, respectively.
The LED reference signals are transmitted on mutually orthogonal
modulation subcarriers and separated at the receiver, while the
RF and optical links are frequency-separated. Accordingly, interference among the LED signals and between the RF and VL links is neglected.

The \(m\)-th LED and the RF node are ceiling-mounted at a common height
\(H\). Their two-dimensional horizontal coordinates are given as
\(\mathbf p_m^{\mathrm{LED}}
=[C_{x,m}^{\mathrm{LED}},C_{y,m}^{\mathrm{LED}}]^T\)
and
\(\mathbf p^{\mathrm{RF}}
=[C_x^{\mathrm{RF}},C_y^{\mathrm{RF}}]^T\),
respectively. Accordingly, their three-dimensional positions are given
by
\(\mathbf{\tilde{p}}_m^{\mathrm{LED}}
=[(\mathbf p^{\mathrm{LED}}_m)^T,H]^T\)
and
\(\mathbf{\tilde{p}}^{\mathrm{RF}}
=[(\mathbf p^{\mathrm{RF}})^T,H]^T\). 
Unlike the fixed LED and RF anchors, the UE moves horizontally while
maintaining a known constant height \(H^{\mathrm{UE}}\), requiring
continuous position and velocity tracking. To this end, the observation period of duration
\(T\) is divided into \(T_{\mathrm{slot}}=T/\Delta t\) time slots
of duration \(\Delta t\), indexed by
\(t=1,\ldots,T_{\mathrm{slot}}\).
The slot duration is assumed to be sufficiently short so that the UE motion and the corresponding RF and VL channels can be regarded as 
constant within each slot. Accordingly, the UE position and velocity
at time slot \(t\) are represented as
\(\widetilde{\mathbf p}_t
=[C_{x,t},C_{y,t},H^{\mathrm{UE}}]^T\)
and
\(\widetilde{\mathbf v}_t
=[v_{x,t},v_{y,t},0]^T\), respectively. The distances from the UE to the \(m\)-th LED and the RF
anchor at time slot \(t\) are respectively given by
\(
d_{m,t}
=
\left\|
\widetilde{\mathbf p}_{t}
-
\widetilde{\mathbf p}_{m}^{\mathrm{LED}}
\right\|_{2}
\)
and
\(
d_t^{\mathrm{RF}}
=
\left\|
\widetilde{\mathbf p}_{t}
-
\widetilde{\mathbf p}^{\mathrm{RF}}
\right\|_{2}.
\)
For notational convenience, the two-dimensional horizontal position and
velocity of the UE are defined as
\(\mathbf p_t=[C_{x,t},C_{y,t}]^T\) and
\(\mathbf v_t=[v_{x,t},v_{y,t}]^T\), respectively. The LEDs are assumed to face vertically downward, while the UE PD
faces vertically upward, with their orientation vectors fixed as
\(\mathbf{o}^{\mathrm{LED}}=[0,0,-1]^T\) and
\(\mathbf{o}^{\mathrm{PD}}=[0,0,1]^T\), respectively. We assume that the LED locations, the RF-anchor location,
and the LED/PD orientation vectors are known to the UE, while the RF link is
perfectly synchronized in time and frequency \cite{VLP-CS, RF_1}.

%To quantify the fundamental gain of RF sensing over conventional VLP, we adopt a simplified model in which 

Based on the geometric and temporal models described above, the
following subsections first derive the VL and RF signal models,
then formulate the UE state-transition model, and finally establish
the recursive Bayesian model underlying joint position and velocity
tracking.
\subsection{VL and RF Signal Models}
For VLP, the UE estimates its position based on the average received optical intensity from the LEDs over each time slot. Since the multipath contribution is typically negligible compared with the dominant LoS component \cite{VLP_Collaborative}, we adopt an LoS-only
optical channel model. After separating the LED reference signals, the VLP observation
from the \(m\)-th LED during time slot \(t\) is expressed as
$y_{m,t}^{\mathrm{VLP}}
=
\sqrt{P_{m,t}^{\mathrm{LED}}}
h_{m,t}^{\mathrm{VLP}}(\mathbf p_t)x_{m,t}
+
n_{m,t}^{\mathrm{VLP}},
  m\in\{1,\ldots,M\}$,
where \(P_{m,t}^{\mathrm{LED}}\) is the transmit power of the
\(m\)-th LED at time slot $t$,
\(x_{m,t}\) is a unit-power reference symbol, and
\(n_{m,t}^{\mathrm{VLP}}\sim
\mathcal{N}(0,(\sigma^{\mathrm{VLP}})^2)\)
denotes the additive noise after signal separation.
For simplicity, we assume identical LED transmit powers,
\(P_{m,t}^{\mathrm{LED}}=P^{\mathrm{LED}}\), and choose
\(x_{m,t}=1\) for all \(m\) and~\(t\). The LoS VL channel gain between the \(m\)-th LED and the UE during the
\(t\)-th time slot is given by \cite{VLP_Collaborative}
\[
\begin{aligned}
h_{m,t}^{\mathrm{VLP}}\left(\mathbf{p}_{t}\right)
&=
\frac{
A_{\mathrm{R}}F_{\mathrm{R}}C_{\mathrm{R}}
}{
2\pi d_{m,t}^{2}
}
(r+1)
\cos^{r}\left(\phi_{m,t}\right)
\cos\left(\psi_{m,t}\right) \\
&=
\Psi_m\frac{
((
\widetilde{\mathbf{p}}_{t}
-
\widetilde{\mathbf{p}}_{m}^{\mathrm{LED}}
)^{T}
\mathbf{o}^{\mathrm{LED}}
)^{r}
\left(
(
\widetilde{\mathbf{p}}_{m}^{\mathrm{LED}}
-
\widetilde{\mathbf{p}}_{t}
)^{T}
\mathbf{o}^{\mathrm{PD}}
\right)}{
d_{m,t}^{r+3}
}
,
\end{aligned}
\]
where \(A_{\mathrm{R}}\) is the physical area of the PD,
\(F_{\mathrm{R}}\) is the optical filter gain, \(C_{\mathrm{R}}\) is the
optical concentrator gain, \(r\) is the Lambertian order, and
\(\Psi_m=A_{\mathrm{R}}F_{\mathrm{R}}C_{\mathrm{R}}(r+1)/(2\pi)\).
As in the conventional LoS VL channel model~\cite{VLP_Collaborative},
\(\phi_{m,t}\) and \(\psi_{m,t}\) denote the irradiance and incidence
angles, respectively. Their cosine factors are determined by the relative
LED--UE geometry and the LED/PD orientation vectors through normalized
inner products, which yields the vector-form expression in the second
equality. Since the UE receives the VL signals from all \(M\) LEDs, the overall VLP observation at time slot \(t\) is collected into the observation vector \(\mathbf y_t^{\mathrm{VLP}}=[y_{1,t}^{\mathrm{VLP}}, \dots, y_{M,t}^{\mathrm{VLP}}]^T\).

We next consider the RF signal model. Unlike VLP, which uses the
average received optical intensity over each time slot, the RF receiver
collects multiple signal samples within the same time slot for
localization and velocity estimation~\cite{CRLB_projection}. During time slot $t$, the RF receiver collects $N^{\mathrm{RF}}$
signal samples, where $\zeta_n$, $n=1,\ldots,N^{\mathrm{RF}}$,
denotes the sampling instant of the $n$-th sample.
Assuming that the received RF signal is dominated by the LoS component
\cite{RF_1, CRLB_projection}, the noiseless received signal at the $n$-th sampling
instant is given by
$\alpha_t b_{t,n}(\mathbf s_t)
=
\alpha_t
e^{j2\pi f_{\mathrm D,t}(\mathbf s_t)\zeta_n}
b\!\left(\zeta_n-\tau_t(\mathbf p_t)\right)$, 
where $\mathbf s_t=[\mathbf p_t^T,\mathbf v_t^T]^T$ denotes the UE
state, $\alpha_t$ is the complex RF channel gain, and $b(\cdot)$
denotes the known RF waveform.
The propagation delay and Doppler frequency are respectively given by
$\tau_t(\mathbf p_t)
= \frac{d_t^{\mathrm{RF}}}{c_0}$
and
$f_{\mathrm D,t}(\mathbf s_t)
= -\frac{(\mathbf p_t-\mathbf p^{\mathrm{RF}})^T\mathbf v_t}
{\lambda d_t^{\mathrm{RF}}}$,
where $c_0$ and $\lambda$ denote the speed of light and the RF
wavelength, respectively.
For notational simplicity, we henceforth write $\tau_t$ and
$f_{\mathrm D,t}$ with their functional arguments omitted.  Collecting the $N^{\mathrm{RF}}$ samples, the RF observation vector is
expressed as
\begin{align}
\mathbf y_t^{\mathrm{RF}}
=
\alpha_t\mathbf b_t(\mathbf s_t)
+
\mathbf n_t^{\mathrm{RF}},
\label{RF_vector_model}
\end{align}
where
$\mathbf y_t^{\mathrm{RF}}
\triangleq
[y_{t,1}^{\mathrm{RF}},\ldots,
y_{t,N^{\mathrm{RF}}}^{\mathrm{RF}}]^T$,
$\mathbf b_t(\mathbf s_t)
\triangleq
[b_{t,1}(\mathbf s_t),\ldots,
b_{t,N^{\mathrm{RF}}}(\mathbf s_t)]^T$,
and
$\mathbf n_t^{\mathrm{RF}}
\sim
\mathcal{CN}(\mathbf 0,
(\sigma^{\mathrm{RF}})^2\mathbf I)$
denotes the RF receiver noise.
For analytical convenience, we assume that the RF reference
signal has state-independent energy over the observation interval, i.e.,
\(
\mathbf b_t(\mathbf s_t)^H\mathbf b_t(\mathbf s_t)=E_b,
\)
where $E_b$ denotes the RF reference-signal energy.
Finally, the RF and VLP observations collected during the \(t\)-th
time slot are jointly represented as
\(
\mathbf{y}_{t}
=
[(\mathbf{y}_{t}^{\mathrm{RF}})^T,
(\mathbf{y}_{t}^{\mathrm{VLP}})^T]^T.
\)

\subsection{State-Transition and Prediction Model}
While the current observation $\mathbf y_t$ provides information about
the UE state at time slot $t$, the current state is also correlated with the previous state through the UE motion. Exploiting this correlation
allows past observations to contribute to the estimation of the current
state. Accordingly, we model the state evolution as
\(
\mathbf{s}_{t}
=
\mathbf{F}\mathbf{s}_{t-1}
+
\mathbf{q}_{t},
\)
where
 \(\mathbf{F}=
\bigl[\begin{smallmatrix}
\mathbf{I}_{2} & \Delta t\,\mathbf{I}_{2}\\
\mathbf{0}_{2 \times 2} & \mathbf{I}_{2}
\end{smallmatrix}\bigr]\)
describes the nominal UE motion according to a constant-velocity
kinematic model. 
Deviations from the nominal constant-velocity motion, caused by acceleration and other motion perturbations, are captured by the process noise
\(\mathbf q_t
=
[(\mathbf q_t^p)^T,(\mathbf q_t^v)^T]^T
\sim\mathcal N(\mathbf 0,\mathbf Q),
\)
where $\mathbf q_t^p$ and $\mathbf q_t^v$ represent the position and
velocity perturbations, respectively. The process-noise covariance
matrix is modeled as
\(\mathbf Q
=
\operatorname{diag}\!\left(
\sigma_{C_x}^2,
\sigma_{C_y}^2,
\sigma_{v_x}^2,
\sigma_{v_y}^2
\right).
\)
We assume that the process noise is independent across time
slots and of the initial state.
% Although the current observation \(\mathbf y_t\) provides
% instantaneous information about the UE state, the estimation accuracy
% can be further improved by exploiting past observations through the UE
% dynamics. To capture the temporal dependence between consecutive time
% slots, we adopt the following UE state-transition model:
% \[
% \mathbf{s}_{t}
% =
% \mathbf{F}\mathbf{s}_{t-1}
% +
% \mathbf{q}_{t},
% \]
% where
% \(\mathbf{F}=
% \bigl[\begin{smallmatrix}
% \mathbf{I}_{2} & \Delta t\,\mathbf{I}_{2}\\
% \mathbf{0}_{2 \times 2} & \mathbf{I}_{2}
% \end{smallmatrix}\bigr]\)
% is the state-transition matrix associated with the constant-velocity
% kinematic model, and
% \(\mathbf q_t=[(\mathbf q_t^p)^T,(\mathbf q_t^v)^T]^T
% \sim\mathcal N(\mathbf 0,\mathbf Q)\)
% models the process uncertainty caused by unobserved acceleration and
% other motion disturbances. The process-noise covariance matrix is
% given by
% \(
% \mathbf Q
% =
% \operatorname{diag}(
% \sigma_{C_x}^2,
% \sigma_{C_y}^2,
% \sigma_{v_x}^2,
% \sigma_{v_y}^2
% )
% \),
% and the process noise is assumed to be independent across time slots
% and independent of the initial state.

We use the subscript $a|b$ to indicate a quantity associated with time slot $a$ conditioned on observations up to time slot $b$.
In particular, let $\hat{\mathbf s}_{t-1|t-1}$ and
$\mathbf P_{t-1|t-1}$ denote, respectively, the state estimate and
its error covariance at time slot $t-1$ based on
$\mathbf y_{1:t-1}$, where
$\mathbf y_{1:t-1}\triangleq\{\mathbf y_1,\ldots,\mathbf y_{t-1}\}$
and
\(\mathbf P_{t-1|t-1}
\triangleq
\mathbb E\!\left[
(\mathbf s_{t-1}-\hat{\mathbf s}_{t-1|t-1})
(\mathbf s_{t-1}-\hat{\mathbf s}_{t-1|t-1})^T
\mid \mathbf y_{1:t-1}
\right]\).
Using the state-transition model, the state estimate and its error
covariance are propagated to time slot $t$ as
\begin{subequations}
\label{s_t_equation}
\begin{align}
\hat{\mathbf{s}}_{t|t-1}
&=
\mathbf{F}\hat{\mathbf{s}}_{t-1|t-1},
\label{state_prediction}\\
\mathbf{P}_{t|t-1}
&=
\mathbf{F}\mathbf{P}_{t-1|t-1}\mathbf{F}^{T}
+
\mathbf{Q}.
\label{covariance_prediction}
\end{align}
\end{subequations}
Here, $\hat{\mathbf s}_{t|t-1}$ and $\mathbf P_{t|t-1}$ denote
the predicted state estimate and its error covariance, respectively,
prior to incorporating the current observation $\mathbf y_t$.

% Let the subscript \(a|b\) indicate that the variable at time slot
% \(a\) is based on the information available up to time slot \(b\).
% Given the final state estimate and error covariance from the previous
% time slot \(t-1\), denoted by \(\hat{\mathbf{s}}_{t-1|t-1}\)
% and \(\mathbf{P}_{t-1|t-1}\), respectively, with
% \(
% \mathbf P_{t-1|t-1}
% \triangleq
% \mathbb E[
% (\mathbf s_{t-1}-\hat{\mathbf s}_{t-1|t-1})
% (\mathbf s_{t-1}-\hat{\mathbf s}_{t-1|t-1})^T
% |\mathbf y_{1:t-1}
% ]
% \),
% the predicted state estimate and error covariance for the current
% time slot \(t\) are obtained using the state-transition model as
% \begin{subequations}
% \label{s_t_equation}
% \begin{align}
% \hat{\mathbf{s}}_{t|t-1}
% &=
% \mathbf{F}\hat{\mathbf{s}}_{t-1|t-1},
% \label{state_prediction}\\
% \mathbf{P}_{t|t-1}
% &=
% \mathbf{F}\mathbf{P}_{t-1|t-1}\mathbf{F}^{T}
% +
% \mathbf{Q}.
% \label{covariance_prediction}
% \end{align}
% \end{subequations}

%Because the posterior estimate at slot \(t-1\) recursively incorporates the accumulated observations \(\mathbf{y}_{1:t-1}\), the prediction for slot \(t\) carries information from all previous slots and serves as the prior for estimating \(\mathbf{s}_t\) from the current RF/VLP observation vector \(\mathbf y_t\).
\subsection{Bayesian Observation and Posterior Model}
Combining the observation and state-transition models introduced above,
we formulate the recursive Bayesian estimation of the UE state.
At time slot $t$, the UE state is characterized by the posterior
distribution $p(\mathbf s_t|\mathbf y_{1:t})$
\cite{estimation_theory}. By Bayes' rule, the posterior distribution
is given by \cite{leon2017probability}
\begin{align}
&p(\mathbf s_t|\mathbf y_{1:t})
=
\frac{
p(\mathbf y_t|\mathbf s_t,\mathbf y_{1:t-1})
p(\mathbf s_t|\mathbf y_{1:t-1})
}{
p(\mathbf y_t|\mathbf y_{1:t-1})
}
\nonumber\\
&\hspace{-0.4cm}\propto\!
p(\mathbf y_t|\mathbf s_t)
p(\mathbf s_t|\mathbf y_{1:t\!-\!1})
\!=\!
p(\mathbf y_t^{\mathrm{RF}}|\mathbf s_t)
p(\mathbf y_t^{\mathrm{VLP}}|\mathbf s_t)
p(\mathbf s_t|\mathbf y_{1:t\!-\!1}),
\label{Bayesian_filtering}
\end{align}
where the first proportionality follows because the current observation
is conditionally independent of the past observations given the current
state, i.e.,
$p(\mathbf y_t|\mathbf s_t,\mathbf y_{1:t-1})
=p(\mathbf y_t|\mathbf s_t)$.
The last equality further uses the conditional independence of the RF
and VLP observations given $\mathbf s_t$.

Under the adopted Gaussian noise models, the RF and VLP likelihoods
follow complex Gaussian and Gaussian distributions, respectively.
Since the nonlinear observation models generally lead to a non-Gaussian
posterior, we employ a Gaussian approximation of the predictive state
distribution to enable tractable recursive estimation
\cite{sarkka2023bayesian}:
\begin{subequations}
\label{pdf_model}
\begin{align}
p\left(\mathbf{y}_{t}^{\mathrm{RF}}|\mathbf{s}_{t}\right)
&=
\mathcal{CN}\left(
\mathbf{y}_{t}^{\mathrm{RF}};
\alpha_t \mathbf{b}_t(\mathbf{s}_{t}),
\mathbf{R}_{t}^{\mathrm{RF}}
\right),
\label{RF_vector PDF}
\\
p\left(\mathbf{y}_{t}^{\mathrm{VLP}}|\mathbf{s}_{t}\right)
&=
\mathcal{N}\left(
\mathbf{y}_{t}^{\mathrm{VLP}};
\boldsymbol{\mu}_{t}^{\mathrm{VLP}}(\mathbf{s}_{t}),
\mathbf{R}_{t}^{\mathrm{VLP}}
\right),
\label{VLP_vector PDF}
\\
p\left(\mathbf{s}_{t}|\mathbf{y}_{1:t-1}\right)
&\approx
\mathcal{N}\left(
\mathbf{s}_{t};
\hat{\mathbf{s}}_{t|t-1},
\mathbf{P}_{t|t-1}
\right).
\label{predict PDF}
\end{align}
\end{subequations}
Here,
$\boldsymbol{\mu}_{t}^{\mathrm{VLP}}(\mathbf{s}_{t})
\triangleq
[\mu_{1,t}^{\mathrm{VLP}}(\mathbf{s}_{t}),\ldots,
\mu_{M,t}^{\mathrm{VLP}}(\mathbf{s}_{t})]^T$,
where
\(\mu_{m,t}^{\mathrm{VLP}}(\mathbf{s}_{t})
\triangleq
\sqrt{P^{\mathrm{LED}}}
h_{m,t}^{\mathrm{VLP}}(\mathbf p_t)x_{m,t}.
\)
The RF and VLP noise covariance matrices are
$\mathbf R_t^{\mathrm{RF}}
=(\sigma^{\mathrm{RF}})^2\mathbf I_{N^{\mathrm{RF}}}$
and
$\mathbf R_t^{\mathrm{VLP}}
=(\sigma^{\mathrm{VLP}})^2\mathbf I_M$,
respectively.
In \eqref{predict PDF}, the predictive distribution of the current
state conditioned on the past observations $\mathbf y_{1:t-1}$ is
approximated as a Gaussian distribution with mean
$\hat{\mathbf s}_{t|t-1}$ and covariance $\mathbf P_{t|t-1}$,
which are recursively computed from the previous state estimate and
its error covariance according to \eqref{s_t_equation}.
Hence, the past observations enter the current Bayesian estimation
through the predictive distribution in \eqref{predict PDF}.

\section{Proposed Algorithm}
Using the Bayesian formulation in Section~\ref{sec:model}, we estimate the current
UE state ${\mathbf s}_{t}$ via MAP
estimation. We formulate the resulting MAP estimation problem as a
nonlinear least-squares (LS) problem and solve it using a damped
Gauss--Newton algorithm \cite{Gauss-newton}.
\subsection{Optimization Problem}
We adopt the MSE as the performance criterion, which is minimized
by the minimum mean-squared error (MMSE) estimator given by the posterior mean. Since direct computation of
the posterior mean is generally intractable, we use the MAP estimate
as a tractable approximation, motivated by a local Gaussian
approximation of the posterior \cite{sarkka2023bayesian}. Accordingly, the
UE state is estimated by solving the following MAP problem:

\begin{align}
\label{MAP_Problem}
&\hat{\mathbf{s}}_{t|t}=\underset{\mathbf{s}_{t}}{\arg\max}\;
p\left(\mathbf{s}_{t}|\mathbf{y}_{1:t}\right)
\nonumber\\
&\overset{(a)}{=}\!
\underset{\mathbf{s}_{t}}{\arg\max}\;
p\left(\mathbf{y}_{t}^{\mathrm{RF}}|\mathbf{s}_{t}\right)
p\left(\mathbf{y}_{t}^{\mathrm{VLP}}|\mathbf{s}_{t}\right)
p\left(\mathbf{s}_{t}|\mathbf{y}_{1:t-1}\right)
\nonumber\\
&\overset{(b)}{=}\!
\underset{\mathbf{s}_{t}}{\arg\min}\;
\Bigg\{
\left\|
\mathbf{r}_{t}^{\mathrm{RF}}(\mathbf{s}_{t})
\right\|_{\left(\mathbf{R}_{t}^{\mathrm{RF}}\right)^{-1}}^{2}
+
\frac{1}{2}
\left\|
\mathbf{r}_{t}^{\mathrm{VLP}}(\mathbf{s}_{t})
\right\|_{\left(\mathbf{R}_{t}^{\mathrm{VLP}}\right)^{-1}}^{2}
\nonumber\\
&\hspace{12mm}
+
\frac{1}{2}
\left\|
\mathbf{r}_{t}^{\mathrm{pre}}(\mathbf{s}_{t})
\right\|_{\mathbf{P}_{t|t-1}^{-1}}^{2}
\Bigg\},
\end{align}
where
\(
\mathbf{r}_{t}^{\mathrm{RF}}(\mathbf{s}_{t})
\triangleq
\mathbf{y}_{t}^{\mathrm{RF}}
-
\hat{\alpha}_{t}
\mathbf{b}_{t}(\mathbf{s}_{t}),
\)
\(
\mathbf{r}_{t}^{\mathrm{VLP}}(\mathbf{s}_{t})
\triangleq
\mathbf{y}_{t}^{\mathrm{VLP}}
-
\pmb{\mu}_{t}^{\mathrm{VLP}}(\mathbf{s}_{t}),
\)  and 
\(
\mathbf{r}_{t}^{\mathrm{pre}}(\mathbf{s}_{t})
\triangleq
\mathbf{s}_{t}
-
\hat{\mathbf{s}}_{t|t-1}
\) denote the RF, VLP, and prediction residuals, respectively, 
and 
\(
\|\mathbf a\|_{\mathbf W}^{2}
\triangleq
\mathbf a^{H}\mathbf W\mathbf a
\)
is the weighted squared norm. Here, (a) follows from \eqref{Bayesian_filtering}, and
(b) follows by taking the negative logarithm of the
Gaussian models in \eqref{RF_vector PDF}--\eqref{predict PDF},
with terms independent of $\mathbf{s}_t$ omitted.

\subsection{Gauss--Newton-Based Tracking Algorithm}
Although \eqref{MAP_Problem} is a weighted LS problem, its RF and VLP
residuals depend nonlinearly on the UE state. We therefore solve
\eqref{MAP_Problem} using a Gauss--Newton-based recursive tracking
algorithm \cite{Gauss-newton}. At each time slot $t$, the algorithm
consists of prediction, state update, and covariance
update. First, the previous state estimate and covariance are
propagated to obtain $\hat{\mathbf s}_{t|t-1}$ and
$\mathbf P_{t|t-1}$. Next, starting from
$\hat{\mathbf s}_{t|t-1}$, Gauss--Newton iterations incorporate the
current RF and VLP observations to obtain $\hat{\mathbf s}_{t|t}$.
Finally, the posterior covariance $\mathbf P_{t|t}$ is computed at
$\hat{\mathbf s}_{t|t}$, and the procedure is repeated at the next
time slot. The detailed procedure is described below.

At each time slot $t$, the predicted state estimate and covariance are
first obtained according to \eqref{s_t_equation}. The Gauss--Newton
iterations are initialized with the predicted state estimate as
\(\mathbf s_t^{(0)}=\hat{\mathbf s}_{t|t-1},
\)
where $\mathbf s_t^{(l)}$ denotes the state estimate at iteration $l$.
We then define the RF and VLP Jacobian matrices at the current iterate
$\mathbf s_t^{(l)}$ as
\begin{subequations}
\label{Jacobians}
\begin{align}
\label{Jacobian_RF}
\mathbf{D}_{t}^{\mathrm{RF},(l)}
&\triangleq
-
\left.
\begin{bmatrix}
\dfrac{\partial\mathbf{r}_{t}^{\mathrm{RF}}}{\partial C_{x,t}}
&
\dfrac{\partial\mathbf{r}_{t}^{\mathrm{RF}}}{\partial C_{y,t}}
&
\dfrac{\partial\mathbf{r}_{t}^{\mathrm{RF}}}{\partial v_{x,t}}
&
\dfrac{\partial\mathbf{r}_{t}^{\mathrm{RF}}}{\partial v_{y,t}}
\end{bmatrix}
\right|_{\mathbf{s}_{t}=\mathbf{s}_{t}^{(l)}},
\\
\label{Jacobian_VLP}
\mathbf{D}_{t}^{\mathrm{VLP},(l)}
&\triangleq
-
\left.
\begin{bmatrix}
\dfrac{\partial\mathbf{r}_{t}^{\mathrm{VLP}}}{\partial C_{x,t}}
&
\dfrac{\partial\mathbf{r}_{t}^{\mathrm{VLP}}}{\partial C_{y,t}}
&
\dfrac{\partial\mathbf{r}_{t}^{\mathrm{VLP}}}{\partial v_{x,t}}
&
\dfrac{\partial\mathbf{r}_{t}^{\mathrm{VLP}}}{\partial v_{y,t}}
\end{bmatrix}
\right|_{\mathbf{s}_{t}=\mathbf{s}_{t}^{(l)}}.
\end{align}
\end{subequations}
We first consider the RF Jacobian
$\mathbf D_t^{\mathrm{RF},(l)}$. Since the RF residual depends on the
unknown channel coefficient $\alpha_t$, we express the LS
estimate of $\alpha_t$ as a function of $\mathbf s_t$:
\begin{align}
\hat{\alpha}_{t}(\mathbf{s}_{t})
&=
\underset{\alpha}{\arg\min}\;
\left\|
\mathbf{y}_{t}^{\mathrm{RF}}
-
\alpha\mathbf{b}_{t}(\mathbf{s}_{t})
\right\|^{2}
=\frac{
\mathbf{b}_{t}(\mathbf{s}_{t})^{H}
\mathbf{y}_{t}^{\mathrm{RF}}
}{
E_b
}.
\label{alpha_function}
\end{align}
Substituting \(\hat{\alpha}_t(\mathbf s_t)\) from 
\eqref{alpha_function} into the RF residual $\mathbf{r}_{t}^{\mathrm{RF}}$ and differentiating with
respect to the \(q\)-th element \(s_q\) of \(\mathbf s_t\) yields
\begin{align}
&\frac{\partial \mathbf r_t^{\mathrm{RF}}}{\partial s_q}
=
-
\frac{\partial\hat{\alpha}_t(\mathbf s_t)}
{\partial s_q}
\mathbf b_t(\mathbf s_t)
-
\hat{\alpha}_t(\mathbf s_t)
\frac{\partial\mathbf b_t(\mathbf s_t)}
{\partial s_q}
\nonumber\\
&=
-
\frac{
\left(
\frac{\partial\mathbf b_t(\mathbf s_t)}
{\partial s_q}
\right)^H
\mathbf y_t^{\mathrm{RF}}
}{
E_b
}
\mathbf b_t(\mathbf s_t)
-
\frac{
\mathbf b_t(\mathbf s_t)^H
\mathbf y_t^{\mathrm{RF}}
}{
E_b
}
\frac{\partial\mathbf b_t(\mathbf s_t)}
{\partial s_q}.
\label{RF_derivate}
\end{align}
The partial derivative
\(
\partial\mathbf b_t(\mathbf s_t)/\partial s_q
\)
in \eqref{RF_derivate} is specified for each state element in
\eqref{RF_channel_derivate}, where
\(\mathbf e_1\) and \(\mathbf e_2\) denote the standard basis vectors
in \(\mathbb R^2\),
\(\mathbf u_t^{\mathrm{RF}}
\triangleq
(\mathbf p_t-\mathbf p^{\mathrm{RF}})/d_t^{\mathrm{RF}}
=[u_{x,t}^{\mathrm{RF}},u_{y,t}^{\mathrm{RF}}]^T\)
is the horizontal projection of the unit direction vector from the
RF anchor to the UE, and
\(\dot b(\xi)\triangleq d b(\xi)/d\xi\)
denotes the derivative of the RF reference waveform with respect to
its argument.
Evaluating \eqref{RF_derivate} using the derivatives in
\eqref{RF_channel_derivate} at
\(\mathbf s_t=\mathbf s_t^{(l)}\) and stacking the resulting
derivatives according to \eqref{Jacobian_RF} yields the RF Jacobian
\(\mathbf D_t^{\mathrm{RF},(l)}\).
\begin{figure*}[t]
\begin{subequations}
\label{RF_channel_derivate}
\begin{align}
\frac{\partial b_{t,n}}{\partial C_{x,t}}
&=
e^{j2\pi f_{D,t}\zeta_n}
\Bigg[
-\frac{j2\pi\zeta_n}{\lambda d_{t}^{\mathrm{RF}}}
b\left(\zeta_n-\tau_{t}\right)
\mathbf{e}_{1}^{T}
\left(
\mathbf{I}_{2}
-
\mathbf{u}_{t}^{\mathrm{RF}}
(\mathbf{u}_{t}^{\mathrm{RF}})^{T}
\right)
\mathbf{v}_{t}
-
\frac{u_{x,t}^{\mathrm{RF}}}{c_0}
\dot{b}
\left(\zeta_n-\tau_{t}\right)
\Bigg], \label{RF_x_derivate}\\
\frac{\partial b_{t,n}}{\partial C_{y,t}}
&=
e^{j2\pi f_{D,t}\zeta_n}
\Bigg[
-\frac{j2\pi\zeta_n}{\lambda d_{t}^{\mathrm{RF}}}
b\left(\zeta_n-\tau_{t}\right)
\mathbf{e}_{2}^{T}
\left(
\mathbf{I}_{2}
-
\mathbf{u}_{t}^{\mathrm{RF}}
(\mathbf{u}_{t}^{\mathrm{RF}})^{T}
\right)
\mathbf{v}_{t} -
\frac{u_{y,t}^{\mathrm{RF}}}{c_0}
\dot{b}
\left(\zeta_n-\tau_{t}\right)
\Bigg], \label{RF_y_derivate} \\
\frac{\partial b_{t,n}}{\partial v_{x,t}}
&=
-\frac{j2\pi\zeta_n u_{x,t}^{\mathrm{RF}}}{\lambda}
e^{j2\pi f_{D,t}\zeta_n}
b\left(\zeta_n-\tau_{t}\right), \quad
\frac{\partial b_{t,n}}{\partial v_{y,t}}
=
-\frac{j2\pi\zeta_n u_{y,t}^{\mathrm{RF}}}{\lambda}
e^{j2\pi f_{D,t}\zeta_n}
b\left(\zeta_n-\tau_{t}\right). \label{RF_v_derivate}
\end{align}
\end{subequations}
\hrulefill
\end{figure*}

Unlike the RF residual, the VLP residual
$\mathbf r_t^{\mathrm{VLP}}(\mathbf s_t)
=
\mathbf y_t^{\mathrm{VLP}}
-
\boldsymbol{\mu}_t^{\mathrm{VLP}}(\mathbf s_t)$
does not involve an unknown channel coefficient. Since the VL channel gain depends only on the UE position,
the derivative of the $m$-th VLP residual element is given by

% \vspace{0.2in}
% Unlike the RF residual, the VLP residual does not involve an
% unknown channel coefficient and is given by
% \(
% \mathbf r_t^{\mathrm{VLP}}(\mathbf s_t)
% =
% \mathbf y_t^{\mathrm{VLP}}
% -
% \boldsymbol{\mu}_t^{\mathrm{VLP}}(\mathbf s_t)
% \),
% whose \(m\)-th element is
% \(
% r_{m,t}^{\mathrm{VLP}}(\mathbf s_t)
% =
% y_{m,t}^{\mathrm{VLP}}
% -
% \mu_{m,t}^{\mathrm{VLP}}(\mathbf s_t)
% \).
% Since the VL channel depends only on the UE position, i.e.,
% \(
% h_{m,t}^{\mathrm{VLP}}(\mathbf s_t)
% =
% h_{m,t}^{\mathrm{VLP}}(\mathbf p_t)
% \),
% and the observation \(y_{m,t}^{\mathrm{VLP}}\) is held fixed when differentiating with respect to \(\mathbf s_t\), the residual derivative is
% given by
\begin{align}
\frac{\partial r_{m,t}^{\mathrm{VLP}}}{\partial s_q}
&=
-
\frac{\partial \mu_{m,t}^{\mathrm{VLP}}}{\partial s_q}
=
-\sqrt{P^{\mathrm{LED}}}\,x_{m,t}
\frac{\partial h_{m,t}^{\mathrm{VLP}}}{\partial s_q},
\label{VLP_residual_derivative}
\end{align}
with 
\begin{subequations}
\label{VLP_derivate}
\begin{align}
\frac{\partial h_{m,t}^{\mathrm{VLP}}}{\partial C_{x,t}}
&=
-(r+3)
h_{m,t}^{\mathrm{VLP}}
\frac{
C_{x,t}-C_{x,m}^{\mathrm{LED}}
}{
d_{m,t}^{2}
},
\\
\frac{\partial h_{m,t}^{\mathrm{VLP}}}{\partial C_{y,t}}
&=
-(r+3)
h_{m,t}^{\mathrm{VLP}}
\frac{
C_{y,t}-C_{y,m}^{\mathrm{LED}}
}{
d_{m,t}^{2}
},
\\
\frac{\partial h_{m,t}^{\mathrm{VLP}}}{\partial v_{x,t}}
&=0,
\qquad
\frac{\partial h_{m,t}^{\mathrm{VLP}}}{\partial v_{y,t}}
=0.
\end{align}
\end{subequations}
Evaluating \eqref{VLP_residual_derivative} using the channel
derivatives in \eqref{VLP_derivate} at
\(\mathbf s_t=\mathbf s_t^{(l)}\) and stacking the resulting
derivatives over all LEDs according to \eqref{Jacobian_VLP} yields
the VLP Jacobian \(\mathbf D_t^{\mathrm{VLP},(l)}\). 

Using the RF and VLP Jacobian matrices, the state is updated at the
$l$-th Gauss--Newton iteration as
\begin{align}
\mathbf{s}_{t}^{(l+1)}
=
\mathbf{s}_{t}^{(l)}
+
\eta_t^{(l)}\Delta\mathbf{s}_{t}^{(l)},
\label{GN_update}
\end{align}
where $\eta_t^{(l)}\in(0,1]$ is the step size and the Gauss--Newton
direction $\Delta\mathbf{s}_{t}^{(l)}$ is given by
\begin{align}
\Delta\mathbf{s}_{t}^{(l)}
&=
\Big[
\mathbf{H}_{t}^{\mathrm{RF},(l)}
+
\mathbf{H}_{t}^{\mathrm{VLP},(l)}
+
\mathbf{P}_{t|t-1}^{-1}
\Big]^{-1}
\nonumber\\
& \times
\Big[
\mathbf{g}_{t}^{\mathrm{RF},(l)}
+
\mathbf{g}_{t}^{\mathrm{VLP},(l)}
-
\mathbf{P}_{t|t-1}^{-1}
\big(
\mathbf{s}_{t}^{(l)}
-
\hat{\mathbf{s}}_{t|t-1}
\big)
\Big],
\label{GN_closed_form}
\end{align}
where
$\mathbf{H}_{t}^{\mathrm{RF},(l)}
\triangleq
2\operatorname{Re} \{
(\mathbf{D}_{t}^{\mathrm{RF},(l)})^{H}
(\mathbf{R}_{t}^{\mathrm{RF}})^{-1}
\mathbf{D}_{t}^{\mathrm{RF},(l)}
 \}$, $\mathbf{g}_{t}^{\mathrm{RF},(l)}
\triangleq
2\operatorname{Re}  \{
(\mathbf{D}_{t}^{\mathrm{RF},(l)})^{H}
(\mathbf{R}_{t}^{\mathrm{RF}})^{-1}
\mathbf{r}_{t}^{\mathrm{RF},(l)}
 \}$, $\mathbf{H}_{t}^{\mathrm{VLP},(l)}
\triangleq
(\mathbf{D}_{t}^{\mathrm{VLP},(l)})^{T}
(\mathbf{R}_{t}^{\mathrm{VLP}})^{-1}
\mathbf{D}_{t}^{\mathrm{VLP},(l)}$, and 
$\mathbf{g}_{t}^{\mathrm{VLP},(l)}
\triangleq
(\mathbf{D}_{t}^{\mathrm{VLP},(l)})^{T}
(\mathbf{R}_{t}^{\mathrm{VLP}})^{-1}
\mathbf{r}_{t}^{\mathrm{VLP},(l)}$. 
Here,
$\mathbf{r}_{t}^{\mathrm{RF},(l)}
\triangleq\mathbf{r}_{t}^{\mathrm{RF}}(\mathbf{s}_t^{(l)})$
and
$\mathbf{r}_{t}^{\mathrm{VLP},(l)}
\triangleq\mathbf{r}_{t}^{\mathrm{VLP}}(\mathbf{s}_t^{(l)})$.

The iterations terminate when
$\|\mathbf{s}_{t}^{(l)}-\mathbf{s}_{t}^{(l-1)}\|_{2}\leq\epsilon$
or when the maximum number of iterations $l_{\text{max}}$ is reached, and the final
iterate is taken as $\hat{\mathbf s}_{t|t}$. Under a local Gaussian
approximation of the posterior around $\hat{\mathbf s}_{t|t}$, the
posterior error covariance is approximated by the inverse of the local
Gauss--Newton information matrix \cite{state_estimation}:
\begin{align}
\mathbf{P}_{t|t}
&\approx
\left[
\mathbf{H}_{t}^{\mathrm{RF},\star}
+
\mathbf{H}_{t}^{\mathrm{VLP},\star}
+
\mathbf{P}_{t|t-1}^{-1}
\right]^{-1},
\label{posterior_covariance}
\end{align}
where
\(\mathbf H_t^{\mathrm{RF},\star}
\triangleq
2\operatorname{Re}\{
(\mathbf D_t^{\mathrm{RF},\star})^H
(\mathbf R_t^{\mathrm{RF}})^{-1}
\mathbf D_t^{\mathrm{RF},\star}
\}\)
and
\(\mathbf H_t^{\mathrm{VLP},\star}
\triangleq
(\mathbf D_t^{\mathrm{VLP},\star})^T
(\mathbf R_t^{\mathrm{VLP}})^{-1}
\mathbf D_t^{\mathrm{VLP},\star}
\)
with $\mathbf D_t^{\mathrm{RF},\star}$ and
$\mathbf D_t^{\mathrm{VLP},\star}$ denoting the corresponding
Jacobians in \eqref{Jacobians} evaluated at
$\mathbf s_t=\hat{\mathbf s}_{t|t}$.
The overall procedure is repeated at each time slot and summarized in
Algorithm~\ref{alg:proposed_algorithm}.

We analyze the computational complexity of the proposed UE state
estimation algorithm at time slot \(t\). Let
\(n_s\triangleq\dim(\mathbf s_t)\) denote the state dimension
(with \(n_s=4\) in the considered system).
The prediction step in \eqref{s_t_equation} requires
\(\mathcal O(n_s^3)\) operations due to the covariance propagation.
In the state-update step, Gauss--Newton iterations are performed up to \(l_{\max}\) times. At each iteration, constructing the RF and VLP
Jacobians requires \(\mathcal O(N^{\mathrm{RF}}n_s)\) and
\(\mathcal O(Mn_s)\) operations, respectively, while forming
\(\mathbf H_t^{\mathrm{RF},(l)}\) and
\(\mathbf H_t^{\mathrm{VLP},(l)}\) requires
\(\mathcal O((N^{\mathrm{RF}}+M)n_s^2)\) operations.
Computing the Gauss--Newton direction in \eqref{GN_closed_form} and
updating the state in \eqref{GN_update} further require
\(\mathcal O(n_s^3)\) operations.
The covariance update in \eqref{posterior_covariance} has the same
order as a single Gauss--Newton iteration. Hence, the overall dominant
complexity per time slot is
\(\mathcal O\!\left(
l_{\max}[(N^{\mathrm{RF}}+M)n_s^2+n_s^3]
\right)\).
Since \(n_s=4\) is fixed, the complexity scales linearly with
\(N^{\mathrm{RF}}+M\) for a fixed \(l_{\max}\). 
\begin{algorithm}[t]
\caption{Proposed Gauss--Newton-Based RF--VLP State Tracking Algorithm}
\label{alg:proposed_algorithm}
\begin{algorithmic}[1]
\STATE \textbf{Input:} \(\hat{\mathbf{s}}_{0|0}\), \(\mathbf{P}_{0|0}\), \(\mathbf{F}\),  \(\mathbf{Q}\), \(\eta_t^{(l)}\in(0,1]\), $l_{\text{max}}$, \(\epsilon\) and RF/VLP observations and system parameters
\FOR{\(t=1,\ldots,T_{\text{slot}}\)}
    \STATE \textbf{Prediction:}
    \(\hat{\mathbf{s}}_{t|t-1}
    =
    \mathbf{F}\hat{\mathbf{s}}_{t-1|t-1}\) and
    \(\mathbf{P}_{t|t-1}
    =
    \mathbf{F}\mathbf{P}_{t-1|t-1}\mathbf{F}^{T}
    +
    \mathbf{Q}\)

    \STATE \textbf{Initialization:}
    \(\mathbf{s}_{t}^{(0)}
    =
    \hat{\mathbf{s}}_{t|t-1}\) and \(l=0\).

    \REPEAT
        \STATE Evaluate
        \(\mathbf{r}_{t}^{\mathrm{RF},(l)}\),
        \(\mathbf{r}_{t}^{\mathrm{VLP},(l)}\),
        \(\mathbf{D}_{t}^{\mathrm{RF},(l)}\), and
        \(\mathbf{D}_{t}^{\mathrm{VLP},(l)}\)
        at \(\mathbf{s}_{t}^{(l)}\) according to
        \eqref{Jacobians}, \eqref{alpha_function}, \eqref{RF_derivate}, \eqref{RF_channel_derivate}, \eqref{VLP_residual_derivative} and
        \eqref{VLP_derivate}

        \STATE Obtain \(\Delta\mathbf{s}_{t}^{(l)}\) using
        \eqref{GN_closed_form}, and update
        \(\mathbf{s}_{t}^{(l+1)}
        =
        \mathbf{s}_{t}^{(l)}
        +
        \eta_t^{(l)}\Delta\mathbf{s}_{t}^{(l)}\)

        \STATE \(l\leftarrow l+1\)

    \UNTIL{
    \(\left\|
    \mathbf{s}_{t}^{(l)}
    -
    \mathbf{s}_{t}^{(l-1)}
    \right\|_{2}
    \leq\epsilon\)
    or \(l=l_{\max}\)}

    \STATE Set
    \(\hat{\mathbf{s}}_{t|t}
    =
    \mathbf{s}_{t}^{(l)}\)

    \STATE Compute \(\mathbf{P}_{t|t}\) using
    \eqref{posterior_covariance}
\ENDFOR
\end{algorithmic}
\end{algorithm}

\section{Bayesian and State-Conditioned Information Analysis}
In this section, we first derive the BIM
and the corresponding BCRLB for the
considered RF--VLP tracking system, which provides a fundamental lower
bound on the Bayesian estimation error. However, the BIM and BCRLB
involve an expectation over the UE state, which makes further
analytical characterization difficult. To enable a tractable analysis, we introduce the SCIM by removing the expectation over the UE
state from the BIM. We further show that, under a first-order
linearization of the observation models, the inverse of the SCIM
coincides with the MSE matrix of the corresponding MAP estimator.
This connection provides an estimation-error interpretation of
the inverse of the SCIM and enables us to characterize the
information and estimation-error behavior of the considered
RF--VLP system.

% In this section, we first derive the Bayesian information matrix (BIM)
% and the corresponding Bayesian Cramér--Rao lower bound (BCRLB) for the
% considered RF--VLP tracking system to establish a Bayesian performance
% reference. However, the BIM and BCRLB involve averaging over the UE state, which makes tractable analytical characterization difficult.
% To obtain tractable analytical insights, we introduce a
% state-conditioned recursive information matrix (SCRM) by removing the
% state averaging from the BIM and define its
% inverse as the state-conditioned inverse-information matrix (SCIM). We further show that, under a first-order linearization of the
% measurement models, the SCIM coincides with the MSE matrix of
% the corresponding MAP estimator. Based on this property, we analyze the SCRM and SCIM to characterize
% the information and estimation-error behavior of the considered RF--VLP system.

\subsection{Information Matrices}
Based on the Bayesian formulation in \eqref{Bayesian_filtering}, the
following theorem derives the BIM and the corresponding BCRLB
\cite{recursive_CRLB}.
\begin{theorem}
\label{theorem:BIM_terms}
The BIM for the UE state
\(\mathbf{s}_{t}\) is given by
\begin{align}
\mathbf{J}_{t}^{\mathrm{B}}
\triangleq
\overline{\mathbf{J}}_{t}^{\mathrm{RF}}
+
\overline{\mathbf{J}}_{t}^{\mathrm{VLP}}
+
\mathbf{J}_{t}^{\mathrm{pred,B}},
\label{Bayesian_information_decomposition}
\end{align}
where the RF, VLP, and prediction information contributions are,
respectively,
\begin{subequations}
\label{BIM_closed_form}
\begin{align}
\overline{\mathbf J}_{t}^{\mathrm{RF}}
&=
\overline{\mathbf J}_{ss,t}^{\mathrm{RF}}
-
\overline{\mathbf J}_{s\alpha,t}^{\mathrm{RF}}
\left(
\overline{\mathbf J}_{\alpha\alpha,t}^{\mathrm{RF}}
\right)^{-1}
\overline{\mathbf J}_{\alpha s,t}^{\mathrm{RF}},
\label{RF_information_matrix}
\\
\overline{\mathbf J}_{t}^{\mathrm{VLP}}
&=
\mathbb E_{\mathbf s_t}
\left[
\sum_{m=1}^{M}
\Gamma_t^{\mathrm{VLP}}
\frac{
(r+3)^{2}
\left(h_{m,t}^{\mathrm{VLP}}\right)^{2}
}{
d_{m,t}^{4}
}
\right.
\nonumber\\
&\qquad\qquad\left.
\times
\begin{bmatrix}
\left(\mathbf p_t-\mathbf p^{\mathrm{LED}}_m\right)
\left(\mathbf p_t-\mathbf p^{\mathrm{LED}}_m\right)^{T}
&
\mathbf{0}_{2\times 2}
\\
\mathbf{0}_{2\times 2}
&
\mathbf{0}_{2\times 2}
\end{bmatrix}
\right],
\label{VLP_information_matrix}
\\
\mathbf{J}_{t}^{\mathrm{pred,B}}
&=
\left[
\mathbf{F}
\left(\mathbf{J}_{t-1}^{\mathrm{B}}\right)^{-1}
\mathbf{F}^{T}
+
\mathbf{Q}
\right]^{-1}.
\label{prediction_information_matrix}
\end{align}
\end{subequations}
Here,
\(\Gamma_t^{\mathrm{VLP}}
\triangleq
P^{\mathrm{LED}}/(\sigma^{\mathrm{VLP}})^{2}\), and
\(
\overline{\mathbf J}_{ab,t}^{\mathrm{RF}}
\triangleq
\mathbb E_{\mathbf s_t}
[
\mathbf J_{ab,t}^{\mathrm{RF}}(\mathbf s_t)
]
\)
for
\(ab\in\{ss,s\alpha,\alpha s,\alpha\alpha\}\),
where
\(\mathbf J_{ab,t}^{\mathrm{RF}}(\mathbf s_t)\)
is given in \eqref{RF_state_joint_FIM_blocks}.
The expectations are taken with respect to the marginal state distribution
of \(\mathbf s_t\).
The corresponding BCRLB is then given by
\begin{align}
\mathbb E\!\left[
(\hat{\mathbf s}_t-\mathbf s_t)
(\hat{\mathbf s}_t-\mathbf s_t)^T
\right]
&\succeq
\left(\mathbf J_t^{\mathrm B}\right)^{-1}.
\label{BCRLB_matrix}
\end{align}
\end{theorem}

\begin{proof}
The detailed derivations of
\(\bar{\mathbf{J}}_{t}^{\mathrm{RF}}\),
\(\bar{\mathbf{J}}_{t}^{\mathrm{VLP}}\), and
\(\mathbf{J}_{t}^{\mathrm{pred,B}}\)
are provided in Appendix~A.
\end{proof}
%Theorem~\ref{theorem:BIM_terms} decomposes the recursive BIM into the RF, VLP, and prediction information terms, thereby allowing the contribution of each component to be analyzed separately.
Although Theorem~\ref{theorem:BIM_terms} provides the BCRLB on the
Bayesian estimation error, the expectation over the UE state in the
BIM makes further analytical characterization difficult. To avoid such
state averaging, several localization studies have considered Fisher
information conditioned on the UE state
\cite{VLP_Distributed_collaborative,VLP_Collaborative,vlp_shadowing}.
Motivated by this approach, we remove the expectation over the UE state
from the BIM and define the SCIM.  
\begin{lemma}
\label{lemma:conditional_information}
For a given UE state trajectory
$\{\mathbf s_{\ell}\}_{\ell=1}^{t}$, the SCIM at time slot $t$ is
given by
\begin{align}
\mathbf J_t^{\mathrm c}
&=
\mathbf J_t^{\mathrm{RF,c}}(\mathbf s_t)
+
\mathbf J_t^{\mathrm{VLP,c}}(\mathbf s_t)
+
\mathbf J_t^{\mathrm{pred,c}},
\label{conditional_information_decomposition}
\end{align}
where
$\mathbf J_t^{\mathrm{RF,c}}(\mathbf s_t)$ and
$\mathbf J_t^{\mathrm{VLP,c}}(\mathbf s_t)$ denote the
state-conditioned information matrices contributed by the RF and VLP
observations, respectively, and
$\mathbf J_t^{\mathrm{pred,c}}$ denotes the information propagated
from the previous time slot. In particular,
$\mathbf J_t^{\mathrm{VLP,c}}(\mathbf s_t)$ is obtained from
\eqref{VLP_information_matrix} without taking the expectation over
$\mathbf s_t$. The RF contribution is given by
\begin{align}
\hspace{-0.2cm}\mathbf J_t^{\mathrm{RF,c}}(\mathbf s_t)
&=
8\pi^{2}\Gamma_t^{\mathrm{RF}}
\mathbf H_{\boldsymbol{\psi},t}
\begin{bmatrix}
\beta^{2}
&
\rho_{\beta T}\beta T_{\mathrm{eff}}
\\
\rho_{\beta T}\beta T_{\mathrm{eff}}
&
T_{\mathrm{eff}}^{2}
\end{bmatrix}
\mathbf H_{\boldsymbol{\psi},t}^{T},
\label{conditional_RF_information}
\end{align}
where
$\Gamma_t^{\mathrm{RF}}
\triangleq
|\alpha_t|^{2}E_b/(\sigma^{\mathrm{RF}})^{2}$
denotes the received RF SNR at time slot $t$,
$\mathbf H_{\boldsymbol{\psi},t}$ is the state-domain
delay--Doppler Jacobian defined in Appendix~A, and
$\beta$, $T_{\mathrm{eff}}$, and $\rho_{\beta T}$ denote the effective
signal bandwidth, effective observation duration, and normalized
delay--Doppler coupling coefficient, respectively, as defined in
Appendix~B. Finally, the prediction-information matrix is recursively obtained from
the SCIM at the previous time slot as
\begin{align}
\mathbf J_t^{\mathrm{pred,c}}
&=
\left[
\mathbf F
\left(\mathbf J_{t-1}^{\mathrm c}\right)^{-1}
\mathbf F^{T}
+
\mathbf Q
\right]^{-1}.
\label{conditioned_prediction_information_matrix}
\end{align}
The recursion is initialized with
$\mathbf J_0^{\mathrm c}=\mathbf P_0^{-1}$,
where $\mathbf P_0$ is the initial state-error covariance.
% The corresponding SCIM is defined as
% \begin{align}
% \mathbf P_t^{\mathrm{c}}
% &\triangleq
% \left(\mathbf J_t^{\mathrm{c}}\right)^{-1},
% \label{SCIM_matrix}
% \end{align}
% with $\mathbf P_0^{\mathrm{c}}=\mathbf P_0$.
 \end{lemma}
\begin{proof}
The VLP term follows directly from
Theorem~\ref{theorem:BIM_terms} by removing the state expectation,
while the prediction term follows from the same recursive relation.
The detailed derivation of
\(\mathbf J_t^{\mathrm{RF,c}}(\mathbf s_t)\)
is provided in Appendix~B.
\end{proof}
We next establish the relationship between the inverse of the SCIM and
the MSE matrix of the MAP estimator using first-order Taylor expansions
of the RF and VLP observation models. Specifically, we expand the
observation models around an operating point
$(\bar{\mathbf s}_t,\bar{\alpha}_t)$.
Let \(\delta\mathbf s_t\triangleq\mathbf s_t-\bar{\mathbf s}_t\),
\(\delta\boldsymbol\alpha_t
\triangleq
[\operatorname{Re}\{\alpha_t-\bar\alpha_t\},
\operatorname{Im}\{\alpha_t-\bar\alpha_t\}]^T\), and
\(\bar{\mathbf b}_t\triangleq\mathbf b_t(\bar{\mathbf s}_t)\). 
We define $\mathbf L_{s,t}^{\mathrm{RF}}$ and
$\mathbf L_{\alpha,t}^{\mathrm{RF}}$ as the Jacobians of
$\alpha\mathbf b_t(\mathbf s)$ with respect to $\mathbf s$ and
$[\operatorname{Re}\{\alpha\},\operatorname{Im}\{\alpha\}]^T$,
respectively, evaluated at
$(\bar{\mathbf s}_t,\bar{\alpha}_t)$. In particular,
\(\mathbf L_{\alpha,t}^{\mathrm{RF}}
=
[\bar{\mathbf b}_t,j\bar{\mathbf b}_t].
\)
Similarly, $\mathbf L_t^{\mathrm{VLP}}$ denotes the Jacobian of
$\boldsymbol\mu_t^{\mathrm{VLP}}(\mathbf s)$ with respect to
$\mathbf s$, evaluated at $\bar{\mathbf s}_t$.
The corresponding first-order Taylor expansions are
\begin{subequations}
\label{eq:local_affine_models}
\begin{align}
\hspace{-0.4cm}\mathbf y_t^{\mathrm{RF}}
\!=&
\bar{\alpha}_t\bar{\mathbf b}_t
\!+\!
\mathbf L_{s,t}^{\mathrm{RF}}\delta\mathbf s_t
\!+\!
\mathbf L_{\alpha,t}^{\mathrm{RF}}\delta\boldsymbol\alpha_t
\!+\!
\mathbf n_t^{\mathrm{RF}}
\!+\!
o\!\left(
\left\|
\begin{bmatrix}
\delta\mathbf s_t\\
\delta\boldsymbol\alpha_t
\end{bmatrix}
\right\|
\right),
\\
\hspace{-0.5cm}\mathbf y_t^{\mathrm{VLP}}
&=
\boldsymbol\mu_t^{\mathrm{VLP}}(\bar{\mathbf s}_t)
+
\mathbf L_t^{\mathrm{VLP}}\delta\mathbf s_t
+
\mathbf n_t^{\mathrm{VLP}}
+
o\!\left(\|\delta\mathbf s_t\|\right), 
\end{align}
\end{subequations}
where $\mathbf n_t^{\mathrm{VLP}}=[n_{1,t}^{\mathrm{VLP}}, n_{2,t}^{\mathrm{VLP}}, ..., n_{M,t}^{\mathrm{VLP}}]^T.$ 
Neglecting the higher-order terms in
\eqref{eq:local_affine_models} yields local affine observation models,
based on which we establish the following proposition.

\begin{proposition}
\label{prop:SCIM_MSE}
At each time slot $t$, consider the local affine RF and VLP observation
models obtained from \eqref{eq:local_affine_models} by neglecting the
higher-order terms, with the Taylor expansions evaluated around
$(\bar{\mathbf s}_t,\bar{\alpha}_t)$. Evaluate the SCIM recursion in
Lemma~\ref{lemma:conditional_information} at the same operating point.
Let $\widehat{\mathbf s}_{t|t}^{\mathrm{lin}}$ denote the MAP estimate
under the resulting first-order model after eliminating the RF nuisance
parameter $\alpha_t$, and define its MSE matrix as
\begin{align}
\mathbf P_t^{\mathrm{lin}}
\triangleq
\mathbb E
\left[
(\widehat{\mathbf s}_{t|t}^{\mathrm{lin}}-\mathbf s_t)
(\widehat{\mathbf s}_{t|t}^{\mathrm{lin}}-\mathbf s_t)^T
\right].
\end{align}
If \(\mathbf P_0^{\mathrm{lin}}=(\mathbf J_0^{\mathrm c})^{-1}\),
then
\begin{align}
\mathbf P_t^{\mathrm{lin}}
=
(\mathbf J_t^{\mathrm c})^{-1}.
\label{MSE_pertubation}
\end{align}
Thus, under the first-order model, the inverse of the SCIM coincides
with the MSE matrix of the MAP estimator.
\end{proposition}

\begin{proof}
We prove the result by induction. The initial equality
\(\mathbf P_0^{\mathrm{lin}}=(\mathbf J_0^{\mathrm c})^{-1}\)
holds by assumption. For the induction step, suppose that
\(\mathbf P_{t-1}^{\mathrm{lin}}
=(\mathbf J_{t-1}^{\mathrm c})^{-1}\).
Together with the common state-transition model and
\eqref{conditioned_prediction_information_matrix}, this implies
\begin{align}
\mathbf P_{t|t-1}^{\mathrm{lin}}
&=
\mathbf F\mathbf P_{t-1}^{\mathrm{lin}}\mathbf F^T+\mathbf Q
=
(\mathbf J_t^{\mathrm{pred,c}})^{-1}.
\label{eq:prediction_covariance_equivalence}
\end{align}

Applying the first-order approximations in
\eqref{eq:local_affine_models} to the Gaussian models underlying
\eqref{MAP_Problem}, before eliminating \(\alpha_t\), yields the
quadratic local MAP objective
\begin{align}
\mathcal L_t^{\mathrm{lin}}
={}&
\left\|
\mathbf y_t^{\mathrm{RF}}
-\bar\alpha_t\bar{\mathbf b}_t
-\mathbf L_{s,t}^{\mathrm{RF}}\delta\mathbf s_t
-\mathbf L_{\alpha,t}^{\mathrm{RF}}\delta\boldsymbol\alpha_t
\right\|_{(\mathbf R_t^{\mathrm{RF}})^{-1}}^2
\nonumber\\
&+
\frac12
\left\|
\mathbf y_t^{\mathrm{VLP}}
-\boldsymbol\mu_t^{\mathrm{VLP}}(\bar{\mathbf s}_t)
-\mathbf L_t^{\mathrm{VLP}}\delta\mathbf s_t
\right\|_{(\mathbf R_t^{\mathrm{VLP}})^{-1}}^2
\nonumber\\
&+
\frac12
\left\|
\delta\mathbf s_t
-(\widehat{\mathbf s}_{t|t-1}^{\mathrm{lin}}
-\bar{\mathbf s}_t)
\right\|_{(\mathbf P_{t|t-1}^{\mathrm{lin}})^{-1}}^2.
\label{eq:local_MAP_objective}
\end{align}
In particular, the RF contribution to the state Hessian satisfies
\begin{align}
\nabla_{\delta\mathbf s_t}^2
\mathcal L_t^{{\mathrm{RF}},{\mathrm{lin}}}
&=
2\operatorname{Re}
\left\{
(\mathbf L_{s,t}^{\mathrm{RF}})^H
(\mathbf R_t^{\mathrm{RF}})^{-1}
\mathbf L_{s,t}^{\mathrm{RF}}
\right\}
=
\mathbf J_{ss,t}^{\mathrm{RF}}(\bar{\mathbf s}_t),
\notag
\end{align}
where \(\mathcal L_t^{{\mathrm{RF}},{\mathrm{lin}}}\) denotes the RF term
in \eqref{eq:local_MAP_objective}, and the second equality follows from
\eqref{Gaussian_RF_FIM} and \eqref{RF_state_joint_FIM_blocks}.
Applying the same procedure to the remaining RF and VLP terms and
including the prediction term yields the joint Hessian
\begin{align}
\boldsymbol\Lambda_t^{\mathrm{lin}}
=
\begin{bmatrix}
(\mathbf P_{t|t-1}^{\mathrm{lin}})^{-1}
+\mathbf J_{ss,t}^{\mathrm{RF}}
+\mathbf J_t^{{\mathrm{VLP}},c}
&
\mathbf J_{s\alpha,t}^{\mathrm{RF}}
\\
\mathbf J_{\alpha s,t}^{\mathrm{RF}}
&
\mathbf J_{\alpha\alpha,t}^{\mathrm{RF}}
\end{bmatrix}.
\label{eq:joint_local_MAP_Hessian}
\end{align}

Since \(\alpha_t\) is not a state variable of interest, it is
eliminated from the local MAP problem. Following the standard
Schur-complement elimination for quadratic least-squares problems
\cite{state_estimation} and using
\eqref{eq:prediction_covariance_equivalence}, the resulting state
Hessian is
\begin{align}
&(\mathbf P_{t|t-1}^{\mathrm{lin}})^{-1}
+
\mathbf J_t^{{\mathrm{VLP}},c}
+
\mathbf J_{ss,t}^{\mathrm{RF}}
-
\mathbf J_{s\alpha,t}^{\mathrm{RF}}
(\mathbf J_{\alpha\alpha,t}^{\mathrm{RF}})^{-1}
\mathbf J_{\alpha s,t}^{\mathrm{RF}}
\nonumber\\
&\quad=
\mathbf J_t^{\mathrm{pred,c}}
+
\mathbf J_t^{{\mathrm{RF,c}}}
+
\mathbf J_t^{{\mathrm{VLP,c}}}
=
\mathbf J_t^{\mathrm c}.
\label{eq:reduced_Hessian_SCRM}
\end{align}
Here, the first equality follows from the RF nuisance elimination,
and the second follows from the SCIM decomposition. All
measurement-information blocks in
\eqref{eq:joint_local_MAP_Hessian} and
\eqref{eq:reduced_Hessian_SCRM} are evaluated at the corresponding
operating points.

For the resulting linear-Gaussian MAP problem, the MSE
matrix of the profile-MAP estimator is given by the inverse of the
reduced state Hessian \cite{state_estimation,estimation_theory}.
Since the reduced state Hessian is \(\mathbf J_t^{\mathrm c}\),
we obtain
\(
\mathbf P_t^{\mathrm{lin}}
=
(\mathbf J_t^{\mathrm c})^{-1}
\),
which establishes \eqref{MSE_pertubation}, closes the induction, and
completes the proof.
\end{proof}

\begin{remark}
\label{remark:local_SCIM}
{For the nonlinear RF--VLP observation models, the inverse
of the SCIM provides a local approximation to the MSE
matrix of the MAP estimator. In the theoretical analysis,
we choose the true state and RF channel coefficient as
the operating point for the first-order linearization
in \eqref{eq:local_affine_models}. This choice ensures that the linearized and
original observation models agree at the true values,
but does not imply that these values are known to the
actual estimator. Since the two models generally differ
away from the operating point, their MAP estimates and
corresponding MSE matrices need not coincide.}
\end{remark}
\subsection{RF-Induced Gain}
In this subsection, we characterize the incremental
information gain and estimation-error reduction provided
by RF sensing at time slot $t$. To distinguish the contribution of the current RF observation from the
effects of past measurements, we compare the SCIM and its inverse with
and without the current RF observation. The prediction information from
previous RF and VLP observations is kept fixed. 
Throughout this subsection, we assume that all required matrix inverses exist and all scalar denominators are nonzero,
and omit the explicit dependence on the UE state for notational simplicity.

First, we partition the SCIM as
\(
\mathbf J_t^{\mathrm c}
=
\begin{bmatrix}
\mathbf A_t & \mathbf C_t\\
\mathbf C_t^{T} & \mathbf D_t
\end{bmatrix},
\)
where \(\mathbf A_t\) and \(\mathbf D_t\) denote the position- and
velocity-information blocks, respectively, and \(\mathbf C_t\)
denotes the cross-information block. For comparison, we define the SCIM excluding the current RF observation as
\(
\boldsymbol{\Lambda}_{0,t}
\triangleq
\mathbf J_t^{\mathrm{VLP,c}}
+
\mathbf J_t^{\mathrm{pred,c}}
=
\begin{bmatrix}
\mathbf A_{0,t} & \mathbf C_{0,t}\\
\mathbf C_{0,t}^{T} & \mathbf D_{0,t}
\end{bmatrix},
\)
where \(\mathbf A_{0,t}\), \(\mathbf C_{0,t}\), and
\(\mathbf D_{0,t}\) denote the corresponding blocks. {We begin by examining the RF-induced position-information gain.}
Taking the Schur complement of the velocity-information block,
we obtain the equivalent position-information matrix
\(
\mathbf E_{\mathbf p,t}
\triangleq
\mathbf A_t-\mathbf C_t\mathbf D_t^{-1}\mathbf C_t^{T}.
\)
Similarly, the equivalent position-information matrix
for the RF-free baseline is
\(
\mathbf E_{\mathbf p,0,t}
\triangleq
\mathbf A_{0,t}
-\mathbf C_{0,t}\mathbf D_{0,t}^{-1}\mathbf C_{0,t}^{T}.
\)
The RF-induced position-SCIM gain is then defined as
\(
\Delta\mathbf J_{\mathbf p,t}^{\mathrm{RF}}
\triangleq
\mathbf E_{\mathbf p,t}-\mathbf E_{\mathbf p,0,t}.
\)
Its closed-form expression and asymptotic behavior are
characterized in the following theorem.

\begin{theorem} \label{theorem:RF_position_gain}
The RF-induced position-SCIM gain at time slot \(t\) is given by
\begin{align}
\Delta\mathbf J_{\mathbf p,t}^{\mathrm{RF}}
&=
\underbrace{
\widetilde{\Gamma}_{t}^{\mathrm{RF}}
\beta^{2}(1-\rho_{\beta T}^{2})
\mathbf r_{\tau,\mathbf p,t}
\mathbf r_{\tau,\mathbf p,t}^{T}
}_{\text{Delay-induced gain}}
\nonumber\\
&\quad+
\underbrace{
\frac{
\widetilde{\Gamma}_{t}^{\mathrm{RF}}T_{\mathrm{eff}}^{2}
}{
1+
\widetilde{\Gamma}_{t}^{\mathrm{RF}}T_{\mathrm{eff}}^{2}
\mathbf r_{f,\mathbf v,t}^{T}
\mathbf D_{0,t}^{-1}
\mathbf r_{f,\mathbf v,t}
}
\mathbf z_t\mathbf z_t^{T}
}_{\text{Doppler-induced gain}},
\label{RF_information_gain}
\end{align}
where
\(\widetilde{\Gamma}_{t}^{\mathrm{RF}}
\triangleq 8\pi^{2}\Gamma_t^{\mathrm{RF}}\);
\(\mathbf r_{\tau,\mathbf p,t}\),
\(\mathbf r_{f,\mathbf p,t}\), and
\(\mathbf r_{f,\mathbf v,t}\) are the delay and Doppler
sensitivity vectors constituting
\(\mathbf H_{\boldsymbol{\psi},t}\), as defined in Appendix~A.
Here,
\(\mathbf z_t
=
\mathbf g_t
-
\mathbf C_{0,t}\mathbf D_{0,t}^{-1}
\mathbf r_{f,\mathbf v,t}\),
with
\(\mathbf g_t
=
\mathbf r_{f,\mathbf p,t}
+
\frac{\rho_{\beta T}\beta}{T_{\mathrm{eff}}}
\mathbf r_{\tau,\mathbf p,t}\).

In the low-SNR regime, as
\(\widetilde{\Gamma}_{t}^{\mathrm{RF}}\rightarrow 0\),
\begin{align}
\Delta\mathbf J_{\mathbf p,t}^{\mathrm{RF}}
&=
\widetilde{\Gamma}_{t}^{\mathrm{RF}}
\Big[
\beta^{2}(1-\rho_{\beta T}^{2})
\mathbf r_{\tau,\mathbf p,t}
\mathbf r_{\tau,\mathbf p,t}^{T}
+
T_{\mathrm{eff}}^{2}
\mathbf z_t\mathbf z_t^{T}
\Big]
\nonumber\\
&\quad+
\mathcal O((\widetilde{\Gamma}_{t}^{\mathrm{RF}})^{2}).
\label{RF_gain_low_SNR}
\end{align}
Thus, both the delay- and Doppler-induced position-SCIM
gains increase linearly with the RF SNR at low SNR.

In contrast, in the high-SNR regime, as
\(\widetilde{\Gamma}_{t}^{\mathrm{RF}}\rightarrow\infty\),
\begin{align}
\Delta\mathbf J_{\mathbf p,t}^{\mathrm{RF}}
&=
\widetilde{\Gamma}_{t}^{\mathrm{RF}}
\beta^{2}(1-\rho_{\beta T}^{2})
\mathbf r_{\tau,\mathbf p,t}
\mathbf r_{\tau,\mathbf p,t}^{T}
\nonumber\\
&\quad+
\frac{
\mathbf z_t\mathbf z_t^{T}
}{
\mathbf r_{f,\mathbf v,t}^{T}
\mathbf D_{0,t}^{-1}
\mathbf r_{f,\mathbf v,t}
}
+
\mathcal O((\widetilde{\Gamma}_{t}^{\mathrm{RF}})^{-1}).
\label{RF_gain_high_SNR}
\end{align}
Thus, the delay-induced position-SCIM gain continues to
increase linearly with the RF SNR, whereas the
Doppler-induced gain saturates at a finite limit.
\end{theorem}
\begin{proof}
The proof is provided in Appendix C.
\end{proof}
We next characterize the RF-induced velocity-information gain.
Taking the Schur complement of the position-information block,
we obtain the equivalent velocity-information matrix
\(
\mathbf E_{\mathbf v,t}
\triangleq
\mathbf D_t
-
\mathbf C_t^{T}\mathbf A_t^{-1}\mathbf C_t.
\)
Similarly, the equivalent velocity-information matrix
for the RF-free baseline is
\(
\mathbf E_{\mathbf v,0,t}
\triangleq
\mathbf D_{0,t}
-
\mathbf C_{0,t}^{T}\mathbf A_{0,t}^{-1}\mathbf C_{0,t}.
\)
The RF-induced velocity-SCIM gain is then defined as
\(
\Delta\mathbf J_{\mathbf v,t}^{\mathrm{RF}}
\triangleq
\mathbf E_{\mathbf v,t}
-
\mathbf E_{\mathbf v,0,t}.
\)
Its closed-form expression and asymptotic behavior are
characterized in the following theorem.

\begin{theorem}
\label{theorem:RF_velocity_gain}
The RF-induced velocity-SCIM gain at time slot \(t\) is given by
\begin{align}
\Delta\mathbf J_{\mathbf v,t}^{\mathrm{RF}}
&=
\underbrace{
\frac{
\widetilde{\Gamma}_{t}^{\mathrm{RF}}\delta_{\tau}
}{
1+\widetilde{\Gamma}_{t}^{\mathrm{RF}}\delta_{\tau}c_t
}
\mathbf w_t\mathbf w_t^{T}
}_{\text{Delay-induced gain}}
\nonumber\\
&\quad+
\underbrace{
\frac{
\widetilde{\Gamma}_{t}^{\mathrm{RF}}T_{\mathrm{eff}}^{2}
}{
1+\widetilde{\Gamma}_{t}^{\mathrm{RF}}T_{\mathrm{eff}}^{2}
b_{v,t}(\widetilde{\Gamma}_{t}^{\mathrm{RF}})
}
\mathbf z_{v,t}(\widetilde{\Gamma}_{t}^{\mathrm{RF}})
\mathbf z_{v,t}^{T}(\widetilde{\Gamma}_{t}^{\mathrm{RF}})
}_{\text{Doppler-induced gain}},
\label{RF_velocity_information_gain}
\end{align}
where
\(\delta_{\tau}
=\beta^{2}(1-\rho_{\beta T}^{2})\),
\(c_t
=\mathbf r_{\tau,\mathbf p,t}^{T}
\mathbf A_{0,t}^{-1}
\mathbf r_{\tau,\mathbf p,t}\),
and
\(\mathbf w_t
=\mathbf C_{0,t}^{T}\mathbf A_{0,t}^{-1}
\mathbf r_{\tau,\mathbf p,t}\).
The Doppler-related quantities are defined as
\(
\mathbf z_{v,t}(\widetilde{\Gamma}_{t}^{\mathrm{RF}})
=
\mathbf r_{f,\mathbf v,t}
-
\mathbf C_{0,t}^{T}
\mathbf A_{\tau,t}^{-1}(\widetilde{\Gamma}_{t}^{\mathrm{RF}})
\mathbf g_t
\)
and
\(
b_{v,t}(\widetilde{\Gamma}_{t}^{\mathrm{RF}})
=
\mathbf g_t^{T}
\mathbf A_{\tau,t}^{-1}(\widetilde{\Gamma}_{t}^{\mathrm{RF}})
\mathbf g_t,
\)
where
\(
\mathbf A_{\tau,t}(\widetilde{\Gamma}_{t}^{\mathrm{RF}})
=
\mathbf A_{0,t}
+
\widetilde{\Gamma}_{t}^{\mathrm{RF}}\delta_{\tau}
\mathbf r_{\tau,\mathbf p,t}
\mathbf r_{\tau,\mathbf p,t}^{T}.
\)

In the low-SNR regime, as
\(\widetilde{\Gamma}_{t}^{\mathrm{RF}}\rightarrow 0\),
\begin{align}
\Delta\mathbf J_{\mathbf v,t}^{\mathrm{RF}}
&\!=\!
\widetilde{\Gamma}_{t}^{\mathrm{RF}}
\Big[
\delta_{\tau}\mathbf w_t\mathbf w_t^{T}
\!+\!
T_{\mathrm{eff}}^{2}
\mathbf z_{v,t}^{0}(\mathbf z_{v,t}^{0})^{T}
\Big]\!+\!
\mathcal O((\widetilde{\Gamma}_{t}^{\mathrm{RF}})^{2}),
\label{RF_velocity_gain_low_SNR}
\end{align}
where \(\mathbf z_{v,t}^{0}\) is the low-SNR limit of
\(\mathbf z_{v,t}(\widetilde{\Gamma}_{t}^{\mathrm{RF}})\),
as explicitly given in Appendix~D.
Thus, both the delay- and Doppler-induced velocity-SCIM
gains increase linearly with the RF SNR at low SNR.

In contrast, in the high-SNR regime, as
\(\widetilde{\Gamma}_{t}^{\mathrm{RF}}\rightarrow\infty\),
\begin{align}
\Delta\mathbf J_{\mathbf v,t}^{\mathrm{RF}}
&=
\frac{1}{c_t}\mathbf w_t\mathbf w_t^{T}
+
\frac{1}{b_{v,t}^{\infty}}
\mathbf z_{v,t}^{\infty}
(\mathbf z_{v,t}^{\infty})^{T}
+
o(1),
\label{RF_velocity_gain_high_SNR}
\end{align}
where \(\mathbf z_{v,t}^{\infty}\) and \(b_{v,t}^{\infty}\)
are the high-SNR limits of
\(\mathbf z_{v,t}(\widetilde{\Gamma}_{t}^{\mathrm{RF}})\)
and \(b_{v,t}(\widetilde{\Gamma}_{t}^{\mathrm{RF}})\),
respectively, as explicitly given in Appendix~D.
Thus, unlike the position-SCIM gain, both the delay-
and Doppler-induced velocity-SCIM gains saturate
at finite limits as the RF SNR increases.
\end{theorem}

\begin{proof}
The detailed derivation of
\(\Delta\mathbf J_{\mathbf v,t}^{\mathrm{RF}}\) and its low- and
high-SNR asymptotic behavior are provided in Appendix~D.
\end{proof}
{The preceding results characterize the RF-induced gains
in position and velocity information. We now examine
how these gains reduce the inverse of the SCIM,
which is closely related to estimation accuracy. } We
partition
\(
(\mathbf J_t^{\mathrm c})^{-1}
=
\begin{bmatrix}
\mathbf X_t & \mathbf S_t\\
\mathbf S_t^{T} & \mathbf V_t
\end{bmatrix}
\)
and
\(
\boldsymbol{\Lambda}_{0,t}^{-1}
=
\begin{bmatrix}
\mathbf X_{0,t} & \mathbf S_{0,t}\\
\mathbf S_{0,t}^{T} & \mathbf V_{0,t}
\end{bmatrix},
\)
where the blocks correspond to position, position--velocity cross,
and velocity components. The RF-induced reduction is defined as
\(
\Delta\mathbf P_t^{\mathrm{RF}}
\triangleq
\boldsymbol{\Lambda}_{0,t}^{-1}
-
(\mathbf J_t^{\mathrm c})^{-1}.
\)
The following theorem gives its closed-form expression and asymptotic
behavior.

\begin{theorem}
\label{Theorem: SCIM_RF_gain}
Define
\(
\mathbf K_{\boldsymbol{\psi}}
\triangleq
\begin{bmatrix}
\beta^{2} &
\rho_{\beta T}\beta T_{\mathrm{eff}}\\
\rho_{\beta T}\beta T_{\mathrm{eff}} &
T_{\mathrm{eff}}^{2}
\end{bmatrix}.
\)
The RF-induced reduction in the inverse of the SCIM
is given by
\begin{align}
\Delta\mathbf P_t^{\mathrm{RF}}
&=
\boldsymbol{\Lambda}_{0,t}^{-1}
\mathbf H_{\boldsymbol{\psi},t}
\left[
(\widetilde{\Gamma}_{t}^{\mathrm{RF}})^{-1}
\mathbf K_{\boldsymbol{\psi}}^{-1}
+
\mathbf H_{\boldsymbol{\psi},t}^{T}
\boldsymbol{\Lambda}_{0,t}^{-1}
\mathbf H_{\boldsymbol{\psi},t}
\right]^{-1}
\nonumber\\
&\quad\times
\mathbf H_{\boldsymbol{\psi},t}^{T}
\boldsymbol{\Lambda}_{0,t}^{-1}.
\label{RF_CRLB_gain}
\end{align}

In the low-SNR regime, as
\(\widetilde{\Gamma}_{t}^{\mathrm{RF}}\rightarrow 0\),
\begin{align}
\Delta\mathbf P_t^{\mathrm{RF}}
&=
\widetilde{\Gamma}_{t}^{\mathrm{RF}}
\boldsymbol{\Lambda}_{0,t}^{-1}
\mathbf H_{\boldsymbol{\psi},t}
\mathbf K_{\boldsymbol{\psi}}
\mathbf H_{\boldsymbol{\psi},t}^{T}
\boldsymbol{\Lambda}_{0,t}^{-1}
+
\mathcal O((\widetilde{\Gamma}_{t}^{\mathrm{RF}})^{2}).
\label{RF_CRLB_low_SNR}
\end{align}
Thus, the RF-induced reduction in the inverse of the SCIM
increases linearly with the RF SNR at low SNR.

In contrast, in the high-SNR regime, as
\(\widetilde{\Gamma}_{t}^{\mathrm{RF}}\rightarrow\infty\),
\begin{align}
\Delta\mathbf P_t^{\mathrm{RF}}
&\rightarrow
\boldsymbol{\Lambda}_{0,t}^{-1}
\mathbf H_{\boldsymbol{\psi},t}
\left[
\mathbf H_{\boldsymbol{\psi},t}^{T}
\boldsymbol{\Lambda}_{0,t}^{-1}
\mathbf H_{\boldsymbol{\psi},t}
\right]^{-1}
\mathbf H_{\boldsymbol{\psi},t}^{T}
\boldsymbol{\Lambda}_{0,t}^{-1}.
\label{RF_CRLB_high_SNR}
\end{align}
Thus, the RF-induced reduction in the inverse of the SCIM
saturates at a finite limit as the RF SNR increases.
\end{theorem}

\begin{proof}
 The detailed derivation of \(\Delta\mathbf P_t^{\mathrm{RF}}\) and its
low- and high-SNR asymptotic behavior are provided in Appendix~E.
\end{proof}
The high-SNR result in Theorem~\ref{Theorem: SCIM_RF_gain} further shows that residual position uncertainty remains even at high RF SNR, as formalized in the following corollary. 
\begin{Corollary}
\label{corollary:RF_position_residual}
{The position block of the inverse of the SCIM remains nonzero as the RF SNR approaches infinity, i.e.,
\begin{align}
\lim_{\widetilde{\Gamma}_{t}^{\mathrm{RF}}\to\infty}
\mathbf X_t
\neq
\mathbf 0_{2\times2}.
\label{RF_position_residual}
\end{align}}
\end{Corollary}
\begin{proof}
From \eqref{RF_CRLB_high_SNR} and
\(
(\mathbf J_t^{\mathrm c})^{-1}
=
\boldsymbol{\Lambda}_{0,t}^{-1}
-
\Delta\mathbf P_t^{\mathrm{RF}}
\),
the high-SNR limit of the inverse of the SCIM is
\begin{align}
\mathbf P_t^{\infty}
&\triangleq
\lim_{\widetilde{\Gamma}_{t}^{\mathrm{RF}}\to\infty}
(\mathbf J_t^{\mathrm c})^{-1}
=
\boldsymbol{\Lambda}_{0,t}^{-1/2}
(\mathbf I_4-\mathbf\Pi_t)
\boldsymbol{\Lambda}_{0,t}^{-1/2},
\end{align}
where
\(\mathbf\Pi_t\triangleq
\boldsymbol{\Lambda}_{0,t}^{-1/2}
\mathbf H_{\boldsymbol{\psi},t}
(\mathbf H_{\boldsymbol{\psi},t}^{T}
\boldsymbol{\Lambda}_{0,t}^{-1}
\mathbf H_{\boldsymbol{\psi},t})^{-1}
\mathbf H_{\boldsymbol{\psi},t}^{T}
\boldsymbol{\Lambda}_{0,t}^{-1/2}\)
is the orthogonal projector onto
\(\operatorname{span}(
\boldsymbol{\Lambda}_{0,t}^{-1/2}
\mathbf H_{\boldsymbol{\psi},t})\).
Thus, \(\mathbf P_t^{\infty}\succeq\mathbf 0\) and
\(
\operatorname{null}(\mathbf P_t^{\infty})
=
\operatorname{span}(\mathbf H_{\boldsymbol{\psi},t})
\).

Let \(\mathbf X_t^{\infty}\) denote the position block of
\(\mathbf P_t^{\infty}\), and suppose that
\(\mathbf X_t^{\infty}=\mathbf 0_{2\times2}\).
For any \(\mathbf u\in\mathbb R^2\), let
\(\bar{\mathbf u}=[\mathbf u^T,\mathbf 0_2^T]^T\).
Then
\(
\bar{\mathbf u}^T\mathbf P_t^{\infty}\bar{\mathbf u}
=
\mathbf u^T\mathbf X_t^{\infty}\mathbf u=0
\),
which, since \(\mathbf P_t^{\infty}\succeq\mathbf 0\), implies
\(
\bar{\mathbf u}\in
\operatorname{span}(\mathbf H_{\boldsymbol{\psi},t})
\)
for every \(\mathbf u\). From \eqref{baseline_information_partition}, this requires scalars
\(a\) and \(b\) such that
\(
[\mathbf u^T,\mathbf 0_2^T]^T
=
a[\mathbf r_{\tau,\mathbf p,t}^T,\mathbf 0_2^T]^T
+
b[\mathbf r_{f,\mathbf p,t}^T,\mathbf r_{f,\mathbf v,t}^T]^T
\).
Since \(\mathbf r_{f,\mathbf v,t}\neq\mathbf 0_2\) for a nondegenerate
RF geometry, the zero velocity component requires \(b=0\), yielding
\(\mathbf u=a\mathbf r_{\tau,\mathbf p,t}\).
This cannot hold for every \(\mathbf u\in\mathbb R^2\).
Therefore,
\(\mathbf X_t^{\infty}\neq\mathbf 0_{2\times2}\). 
\end{proof}
 
\begin{remark}
Theorems~\ref{theorem:RF_position_gain} and~\ref{theorem:RF_velocity_gain}  reveal distinct high-SNR behaviors
of the RF-induced position- and velocity-SCIM gains.
The position-information gain continues to increase
with the RF SNR because delay provides increasingly
precise information along the range direction. For velocity information, the delay-induced gain saturates because velocity is inferred only indirectly from the range through the prediction information. The Doppler-induced gain also saturates because Doppler couples position and velocity, and the residual position uncertainty in Corollary~\ref{corollary:RF_position_residual} limits further velocity-information gain.
\end{remark}
% \begin{remark}
% \cd{Theorems~2 and~\ref{theorem:RF_velocity_gain} reveal distinct high-SNR behaviors of the RF-induced position- and velocity-SCIM gains. The position-SCIM gain continues to increase with the RF SNR, whereas the velocity-SCIM gain saturates at a finite limit. Because delay is directly related to position, increasing the RF SNR
% continues to strengthen the position information. In contrast, Doppler couples position and velocity, so the residual
% position uncertainty established in
% Corollary~\ref{corollary:RF_position_residual} limits further gains in
% velocity information at high RF SNR.}
% \end{remark}
\begin{remark}
As seen from \eqref{RF_CRLB_high_SNR}, the high-SNR limit of the
RF-induced reduction in the inverse of the SCIM depends on the
RF-free baseline uncertainty \(\boldsymbol{\Lambda}_{0,t}^{-1}\), as
well as on the RF measurement sensitivity. Hence, the estimation-accuracy
improvement provided by RF sensing also depends on the available VLP
and prediction information.
\end{remark}

% \begin{remark}
% The reduction in the inverse of the SCIM explicitly depends on \(\boldsymbol{\Lambda}_{0,t}^{-1}\),
% which is determined by the VLP and prediction information. Hence, the
% RF-induced reduction in the inverse of the SCIM depends on the underlying RF-free estimation accuracy.
% \end{remark}
\subsection{VLP-SNR Asymptotic Analysis}
%In the previous subsection, we analyzed the RF-induced gains in the SCRM and SCIM, since multiple VLP observations provide sufficient state information to establish a well-defined baseline. In contrast, with only a single RF link, the RF-only information matrix can be rank-deficient and its inverse may not be well defined. Accordingly, rather than defining a VLP-induced gain relative to an RF-only baseline, we directly characterize the asymptotic behavior of the SCRM and SCIM with respect to the VLP SNR.
% In the previous subsection, we characterized the RF-induced gains
% relative to a baseline consisting of VLP and prediction information.
% In contrast, a single RF link provides only range and radial-velocity information, so the RF-and-prediction baseline can retain substantial state uncertainty and is less informative for isolating a VLP-induced gain. We therefore examine how the SCIM and its inverse vary with the
% VLP SNR, rather than introducing a separate VLP-induced gain metric.
In this subsection, we investigate how the VLP SNR affects the SCIM and its inverse.
Specifically, we vary the VLP SNR at the current time slot $t$ while keeping the RF and prediction
information fixed, thereby isolating the effect of the current VLP observations.

From Lemma~\ref{lemma:conditional_information}, the VLP contribution
to the SCIM provides no direct velocity information, while its
position-information component scales linearly with
\(\Gamma_t^{\mathrm{VLP}}\). Given this explicit dependence on the VLP
SNR, we focus on the asymptotic behavior of the inverse of the SCIM.
For notational convenience, we define the VLP-free SCIM as
\(
\boldsymbol{\Lambda}_t^{\mathrm{RP}}
\triangleq
\mathbf J_t^{\mathrm{RF,c}}
+
\mathbf J_t^{\mathrm{pred,c}}
=
\begin{bmatrix}
\mathbf A_t^{\mathrm{RP}} &
\mathbf C_t^{\mathrm{RP}}\\
(\mathbf C_t^{\mathrm{RP}})^T &
\mathbf D_t^{\mathrm{RP}}
\end{bmatrix}.
\)
The VLP contribution can be expressed as
\(
\mathbf J_t^{\mathrm{VLP,c}}
=
\Gamma_t^{\mathrm{VLP}}
\begin{bmatrix}
\mathbf G_t^{\mathrm{VLP}} & \mathbf 0_{2\times2}\\
\mathbf 0_{2\times2} & \mathbf 0_{2\times2}
\end{bmatrix},
\)
where \(\mathbf G_t^{\mathrm{VLP}}\) denotes the position-information
block normalized by the VLP SNR. The following theorem characterizes
the asymptotic behavior of the inverse of the SCIM as the VLP SNR increases.

\begin{theorem}
\label{theorem:VLP_SCIM_asymptotic}
Assume that
$\boldsymbol{\Lambda}_t^{\mathrm{RP}}$,
$\mathbf D_t^{\mathrm{RP}}$, and
$\mathbf G_t^{\mathrm{VLP}}$
are nonsingular, and define
$\mathbf P_t^{\mathrm{RP}}
\triangleq
(\boldsymbol{\Lambda}_t^{\mathrm{RP}})^{-1}$.

In the low-SNR regime, as
$\Gamma_t^{\mathrm{VLP}}\rightarrow 0$,
\begin{align}
(\mathbf J_t^{\mathrm c})^{-1}
&\!=\!
\mathbf P_t^{\mathrm{RP}}
\!-\!
\Gamma_t^{\mathrm{VLP}}
\mathbf P_t^{\mathrm{RP}}
\!\begin{bmatrix}
\mathbf G_t^{\mathrm{VLP}} & \mathbf 0_{2\times2}\\
\mathbf 0_{2\times2} & \mathbf 0_{2\times2}
\end{bmatrix}\!
\mathbf P_t^{\mathrm{RP}}\!+\!
\mathcal O((\Gamma_t^{\mathrm{VLP}})^2).
\label{VLP_CRLB_low_SNR}
\end{align}
Thus, the reduction in the inverse of the SCIM scales linearly
with the VLP SNR in the low-SNR regime.

In the high-SNR regime, i.e., as
$\Gamma_t^{\mathrm{VLP}}\rightarrow\infty$,
\begin{align}
(\mathbf J_t^{\mathrm c})^{-1}
&=
\begin{bmatrix}
\mathbf 0_{2\times2} & \mathbf 0_{2\times2}\\
\mathbf 0_{2\times2} & (\mathbf D_t^{\mathrm{RP}})^{-1}
\end{bmatrix}
+
\frac{1}{\Gamma_t^{\mathrm{VLP}}}
\begin{bmatrix}
(\mathbf G_t^{\mathrm{VLP}})^{-1} & -\mathbf Z_t\\
-\mathbf Z_t^T & \mathbf W_t
\end{bmatrix}
\notag\\
&\quad+
\mathcal O((\Gamma_t^{\mathrm{VLP}})^{-2}),
\label{VLP_CRLB_high_SNR}
\end{align}
where
$\mathbf Z_t
\triangleq
(\mathbf G_t^{\mathrm{VLP}})^{-1}
\mathbf C_t^{\mathrm{RP}}
(\mathbf D_t^{\mathrm{RP}})^{-1}$
and
$\mathbf W_t
\triangleq
(\mathbf D_t^{\mathrm{RP}})^{-1}
(\mathbf C_t^{\mathrm{RP}})^T
(\mathbf G_t^{\mathrm{VLP}})^{-1}
\mathbf C_t^{\mathrm{RP}}
(\mathbf D_t^{\mathrm{RP}})^{-1}$.
Consequently,
\begin{align}
\mathbf X_t
&\rightarrow \mathbf 0_{2\times2},
&
\mathbf V_t
&\rightarrow(\mathbf D_t^{\mathrm{RP}})^{-1},
\label{VLP_CRLB_limits}
\end{align}
with $\mathbf X_t
=\mathcal O((\Gamma_t^{\mathrm{VLP}})^{-1})$. Thus, as the VLP SNR increases, the position uncertainty
vanishes, whereas the velocity uncertainty approaches
a finite limit determined by the RF and prediction information.
\end{theorem}
\begin{proof}
The proof is provided in Appendix F.
\end{proof}
\begin{remark}
{Corollary~\ref{corollary:RF_position_residual} shows that
the position block $\mathbf X_t$ of the inverse of the
SCIM remains nonzero as the RF SNR approaches infinity,
whereas Theorem~\ref{theorem:VLP_SCIM_asymptotic}
shows that it approaches zero as the VLP SNR increases.
This difference stems from the richer geometric
diversity provided by multiple VL links compared
with the limited directional information available
from a single RF link.}
\end{remark}

%\begin{remark}
%Although VLP does not directly measure velocity, \eqref{VLP_CRLB_exact} shows that increasing the VLP SNR can reduce both position and velocity estimation uncertainty, with the velocity SCIM approaching a finite limit. To further examine this velocity benefit, consider a simplified \(N\)-slot setting with fixed VLP geometry, constant-velocity motion, and negligible process noise. Under these assumptions, the accumulated VLP information matrix for the initial state is
%\(
%\mathbf G_N
%=
%\Gamma_t^{\mathrm{VLP}}
%\sum_{k=0}^{N-1}
%(\mathbf F^k)^T
%\begin{bmatrix}
%\mathbf L&\mathbf 0\\
%\mathbf 0&\mathbf 0
%\end{bmatrix}
%\mathbf F^k
%\)
%\cite{qian2022model,konig2023time}.
%For \(N\geq2\) and nonsingular \(\mathbf L\), eliminating the position parameters through the Schur complement yields the equivalent velocity information
%\begin{align}
%\mathbf G_{\mathbf v|\mathbf p,N}^{\mathrm{VLP}}
%&=
%\Gamma_t^{\mathrm{VLP}}\Delta t^2
%\frac{N(N^2-1)}{12}\mathbf L.
%\label{VLP_temporal_velocity_information}
%\end{align}
%Thus, successive VLP position measurements provide indirect velocity information through the state-transition model.
%\end{remark}
\begin{remark}
Although VLP does not directly measure velocity, successive position
measurements provide indirect velocity information through the
state-transition model. Consequently, increasing the VLP SNR can
reduce the velocity block \(\mathbf V_t\) although it converges to a finite limit.
\end{remark}
\section{Numerical Results}
\begin{figure}[t]
    \centering
    \includegraphics[width=8.1cm]{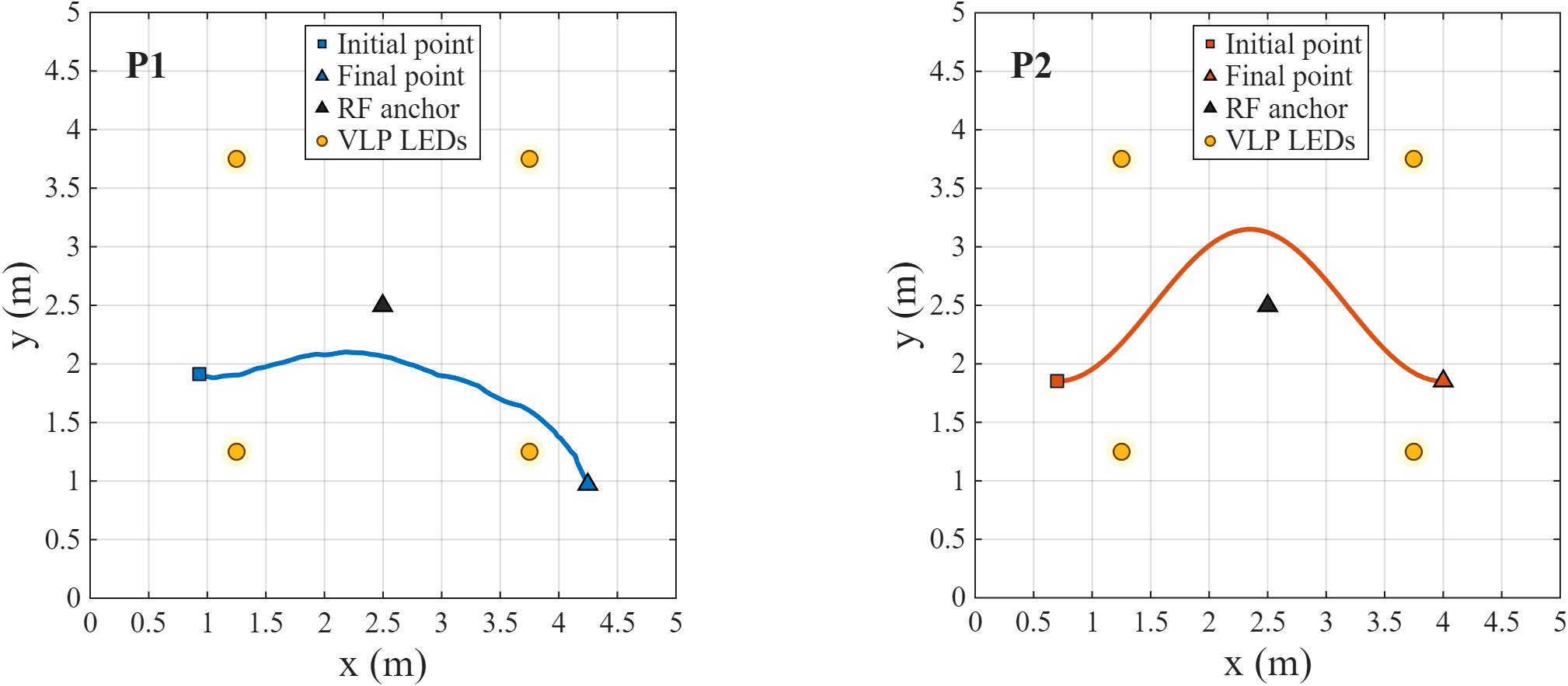}
    \caption{RF--VLP system configuration with Gaussian (P1)
and sinusoidal (P2) UE motion.}
    \label{main path}
\end{figure}
\begin{figure}[t]
    \centering
    \includegraphics[width=8.1cm]{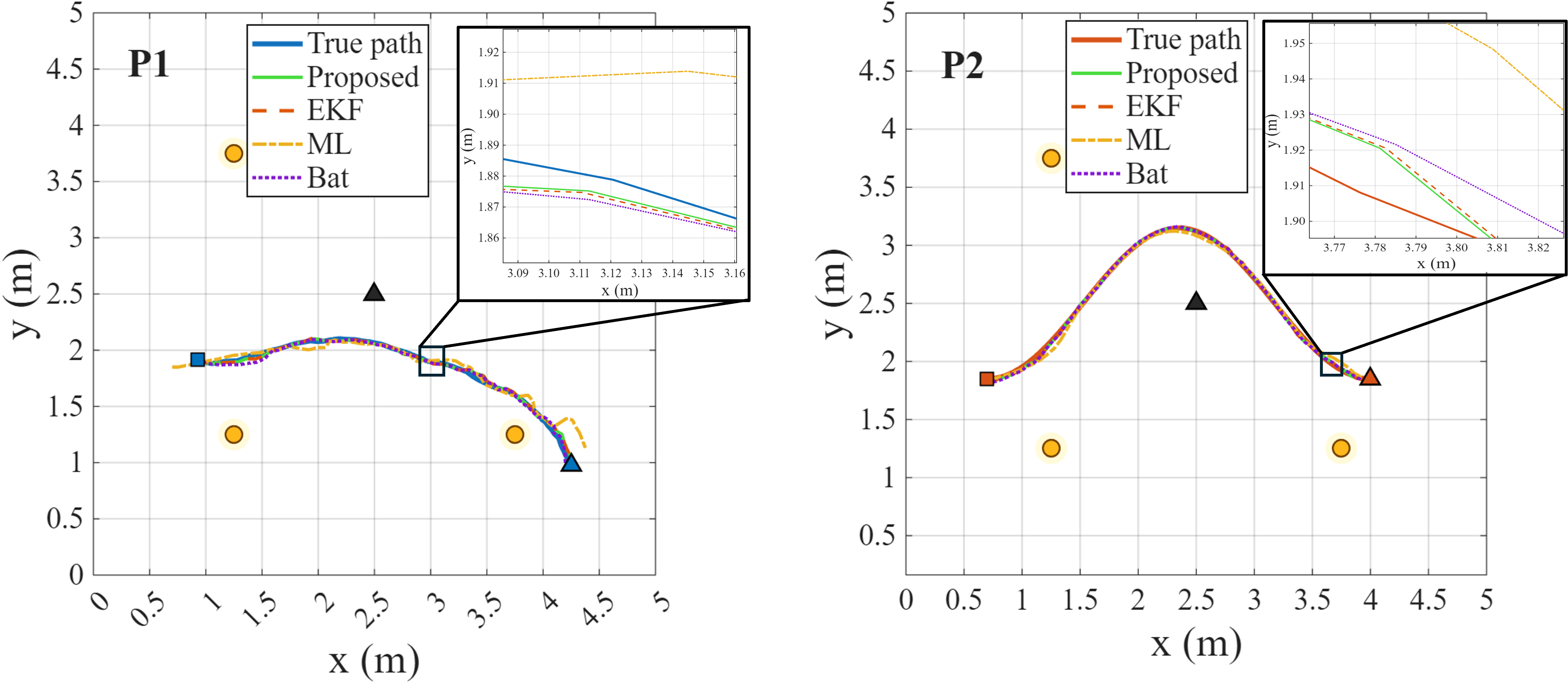}
    \caption{True and estimated UE trajectories for P1 and P2.}
    \label{estimated path}
\end{figure}
\begin{figure}[t]
    \centering
    \includegraphics[width=8.1cm]{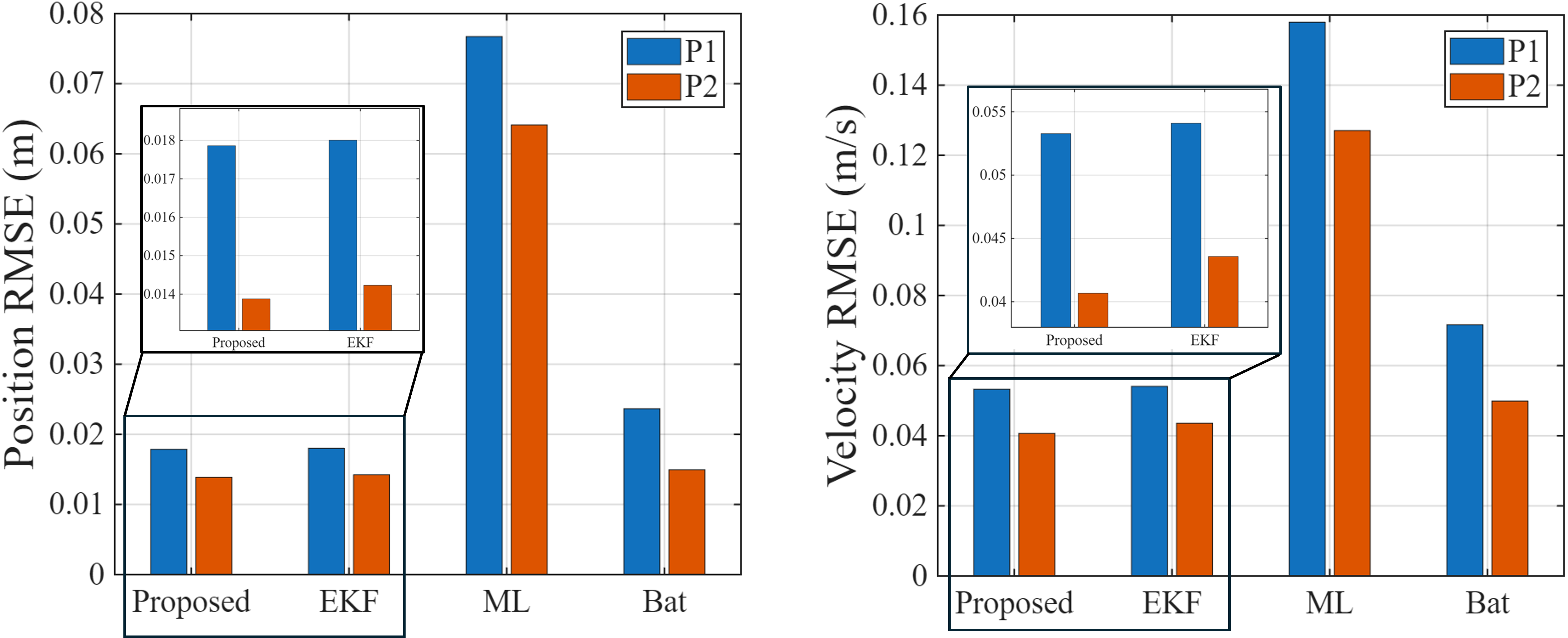}
    \caption{Position and velocity RMSE comparison of the proposed and baseline algorithms for P1 and P2.}
    \label{estimation_value}
\end{figure}
%\begin{figure}[t]
    %\centering
    %\includegraphics[width=8cm]{RMSE_versus_Bayesian_PCRB.png}
    %\caption{RMSE and BCRLB versus transmit power for RF--VLP and 5-%VLP schemes along P1.}
%    \label{rmse vs crlb}
%\end{figure}
\begin{figure}[t]
    \centering
    \includegraphics[width=8cm]{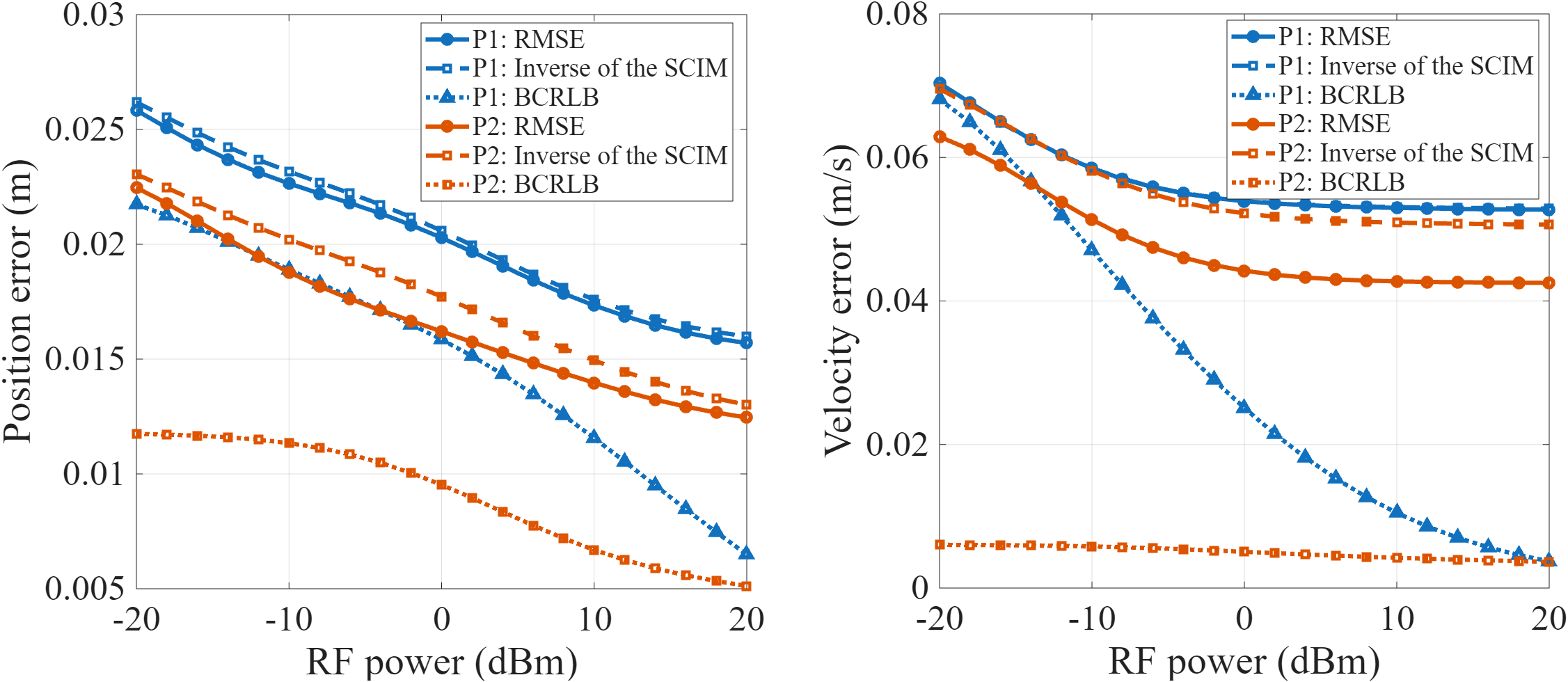}
    \caption{Comparison of position and velocity RMSEs with BCRLB- and inverse-SCIM-based error measures versus RF transmit power for P1 and P2.}
    \label{SCIM vs RMSE}
\end{figure}
\begin{figure}[t]
    \centering
    \includegraphics[width=8.1cm]{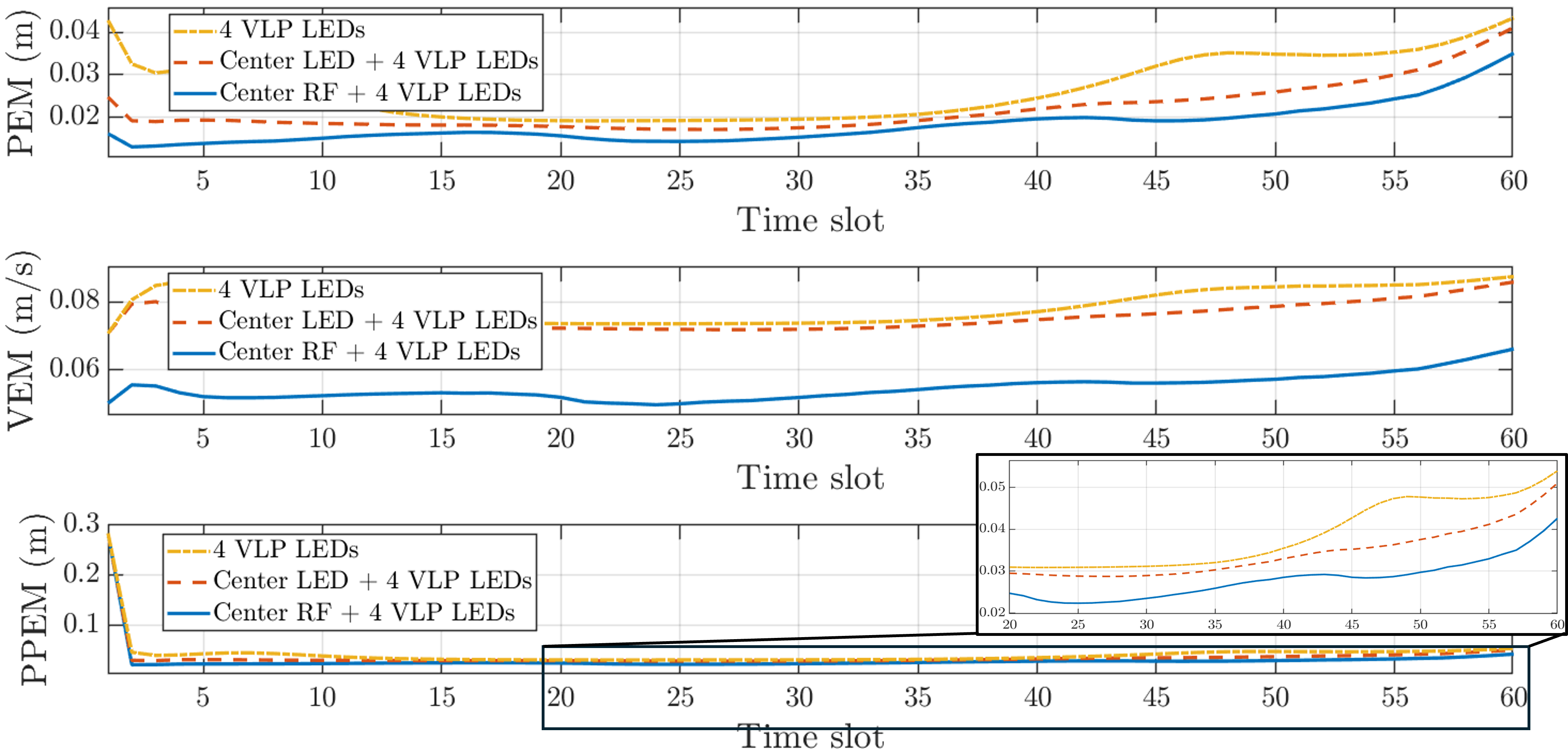}
\caption{Comparison of PEM, VEM, and PPEM versus time slot for RF--VLP and VLP-only configurations along P1.}
    \label{timeslot}
\end{figure}
\begin{figure}[t]
    \centering
    \includegraphics[width=8.1cm]{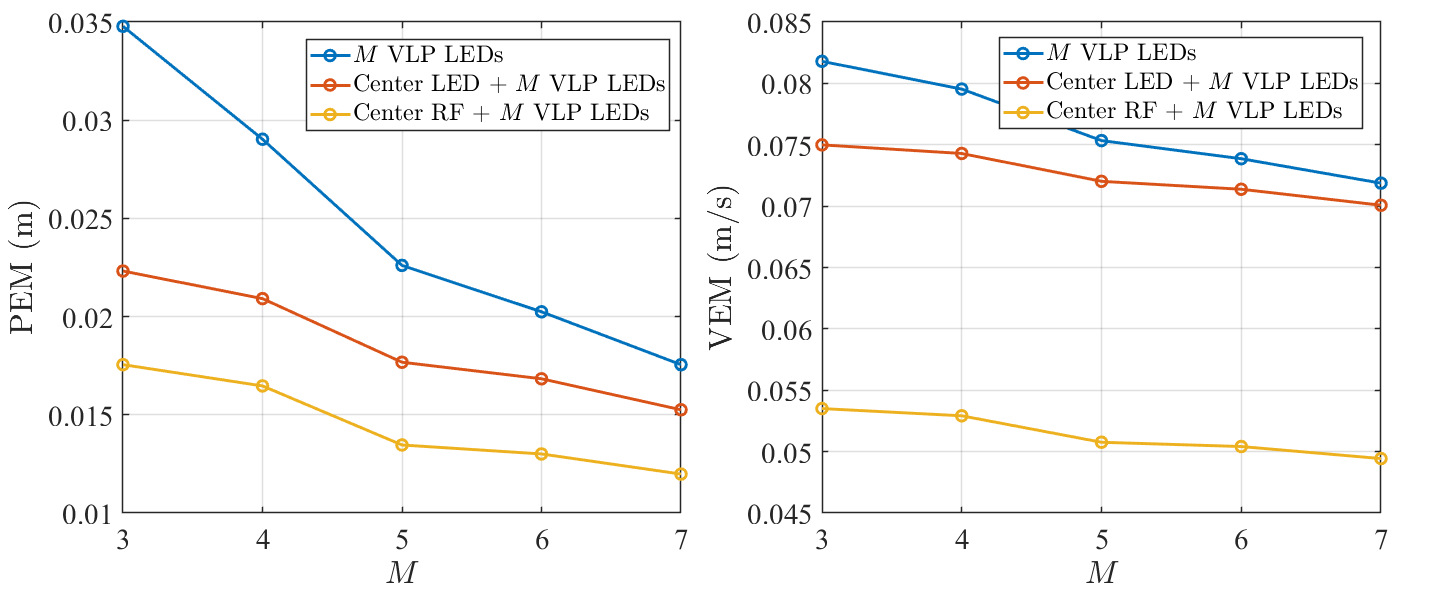}
    \caption{Comparison of PEM and VEM versus the number of LEDs
for RF--VLP and VLP-only configurations along P1.}
    \label{fig:LED_variation}
\end{figure}
\begin{figure}[t]
    \centering
    \includegraphics[width=8.1cm]{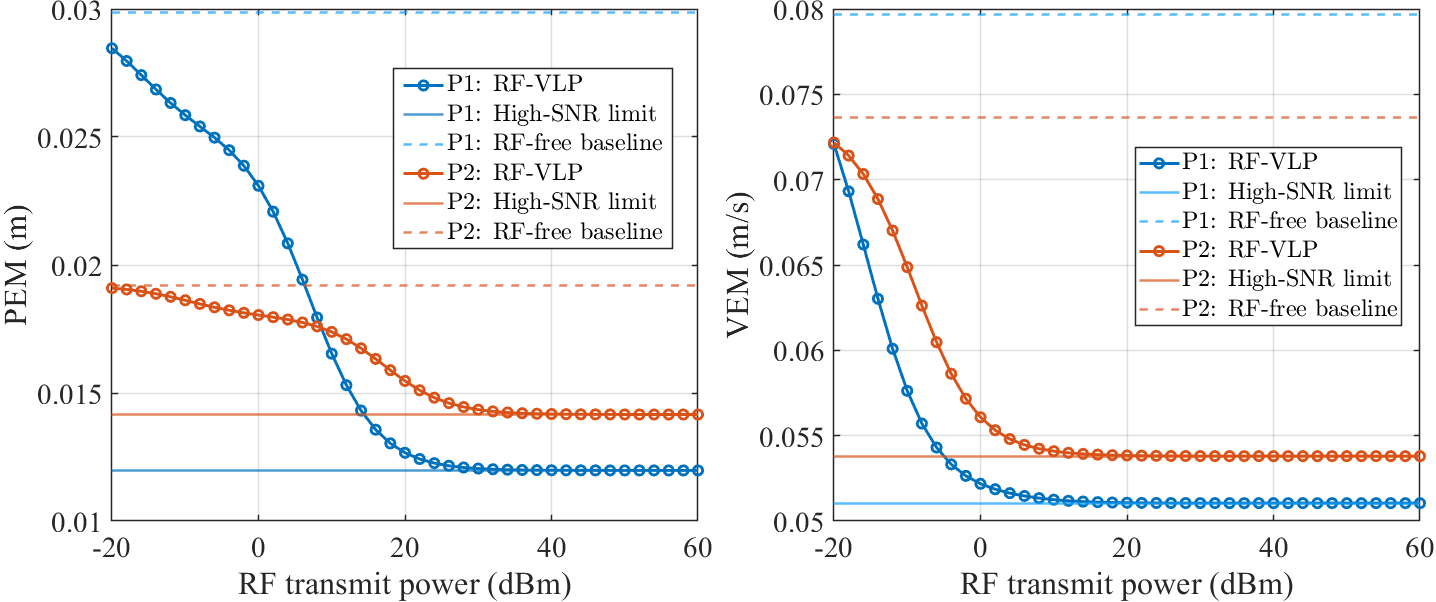}
    \caption{PEM and VEM versus RF transmit power at $t=40$, with RF-free baselines and high-SNR limits.}
    \label{Thorem3_RF variation}
\end{figure}
\begin{figure}[t]
    \centering
    \includegraphics[width=8.1cm]{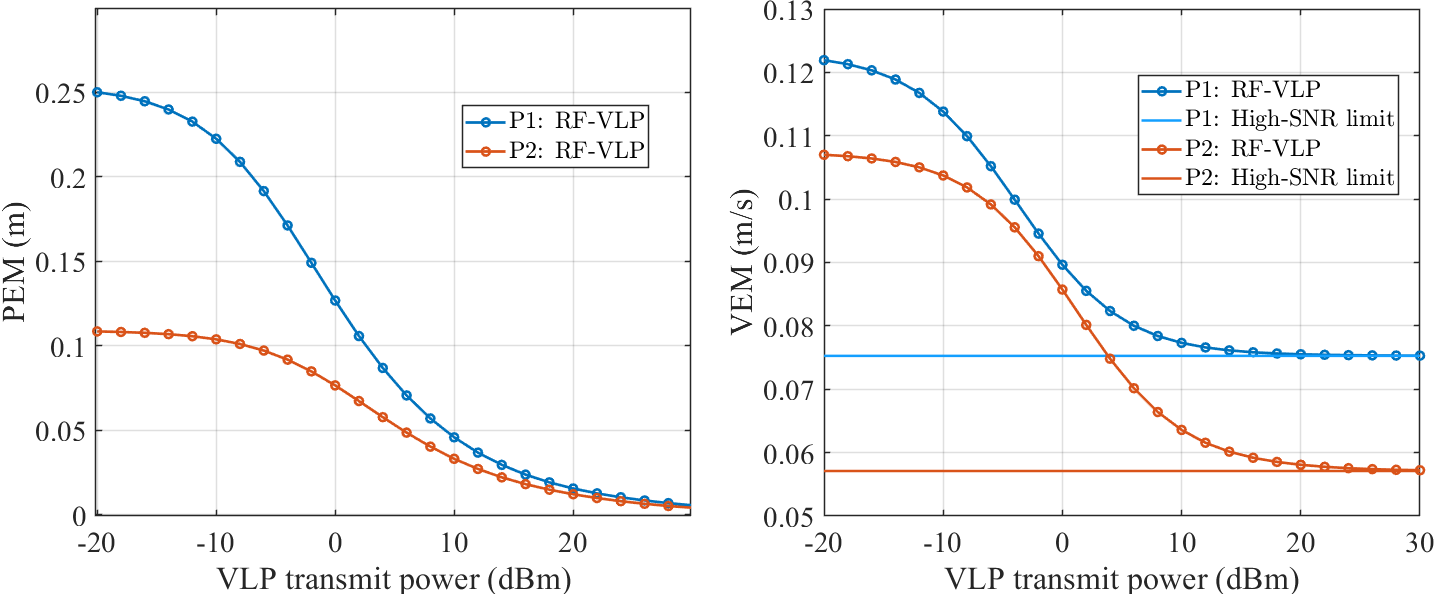}
    \caption{PEM and VEM versus VLP transmit power at $t=40$, with high-SNR limits.}
    \label{Thorem4_VLP variation}
\end{figure}

In this section, we numerically evaluate the proposed algorithm and
validate the derived BCRLB, the SCIM and its inverse, and the corresponding theoretical results.

As shown in Fig.~\ref{main path}, we consider a \(5~\mathrm{m}\times5~\mathrm{m}\)
indoor area with a ceiling height of \(3~\mathrm{m}\), where four LEDs
are deployed near the corners and one RF anchor is located at the
ceiling center. Two UE trajectories are considered: P1 follows the
assumed Gaussian motion model, whereas P2 is a sinusoidal trajectory
used to examine motion-model mismatch. The UE moves at a height of
\(0.85~\mathrm{m}\) with \(\Delta t=0.2~\mathrm{s}\) over
\(T_{\mathrm{slot}}=60\) slots. The initial and process-noise
covariances are set to
\(\mathbf P_0=\mathrm{diag}(0.20^2,0.20^2,0.05^2,0.05^2)\) and
\(\mathbf Q=\mathrm{diag}(5\times10^{-5},5\times10^{-5},
10^{-3},10^{-3})\), respectively. The RF carrier frequency is set to
$f_c=3.5~\mathrm{GHz}$, with $N^{\mathrm{RF}}=32$
samples per time slot.
The default transmit power is set to $15$ dBm
for both the RF anchor and each LED.
The corresponding RF and VLP SNRs are set
to $15$ and $30$ dB, respectively.

To evaluate the estimation performance, we define
the position and velocity RMSEs as
$\sqrt{\mathbb E[
\|\widehat{\mathbf p}_{t|t}-\mathbf p_t\|^2]}$
and
$\sqrt{\mathbb E[
\|\widehat{\mathbf v}_{t|t}-\mathbf v_t\|^2]}$,
respectively, where
$\widehat{\mathbf p}_{t|t}$ and
$\widehat{\mathbf v}_{t|t}$ are the position and
velocity components of $\widehat{\mathbf s}_{t|t}$.
For theoretical performance evaluation, we define
the position error measure (PEM), velocity error
measure (VEM), and prediction-based position error
measure (PPEM) as $\mathrm{PEM}_t\triangleq
\sqrt{[\mathbf P_t]_{1,1}+[\mathbf P_t]_{2,2}}, 
\mathrm{VEM}_t
\triangleq
\sqrt{[\mathbf P_t]_{3,3}+[\mathbf P_t]_{4,4}}$, and $ 
\mathrm{PPEM}_t
\triangleq
\sqrt{
[\mathbf F\mathbf P_{t-1}\mathbf F^T+\mathbf Q]_{1,1}
+
[\mathbf F\mathbf P_{t-1}\mathbf F^T+\mathbf Q]_{2,2}
}$.
Here, $\mathbf P_t=(\mathbf J_t^{\mathrm B})^{-1}$
or $\mathbf P_t=(\mathbf J_t^{\mathrm c})^{-1}$
is used to obtain the corresponding BCRLB-based
or inverse-SCIM-based measures.
Monte Carlo results are averaged over $100$ trials. Unless otherwise specified, the performance measures are additionally averaged over all
$T_{\mathrm{slot}}=60$ time slots.

Figs.~\ref{estimated path} and~\ref{estimation_value}
compare the proposed algorithm with a two-stage
extended Kalman filter (EKF) that uses estimated
RF delay/Doppler and VLP measurement parameters
\cite{hong2026tracking}, a maximum-likelihood (ML)
estimator without predictive prior information
\cite{estimation_theory}, and a Bat-based MAP solver
\cite{VLP-bat}.
Fig.~\ref{estimated path} illustrates the estimated
trajectories, while Fig.~\ref{estimation_value}
compares the corresponding RMSEs.
The proposed algorithm achieves the lowest RMSEs
for both trajectories.
The EKF performs comparably for P1 but shows a
larger performance gap for P2, suggesting that
the proposed algorithm is less sensitive to
motion-model mismatch.
The ML estimator exhibits larger errors
because it does not exploit predictive prior information.

% \textcolor{gray}{Fig.~\ref{rmse vs crlb} compares the RMSE and BCRLB for the RF + 4VLP
% and 5 VLP configurations, where the latter replaces the RF node with a center
% LED using the same transmit power. For BCRLB validation, the state expectations in the BIM are approximated using \(100\) Monte Carlo realizations under the assumed
% Gaussian motion model. The BCRLB remains
% below the RMSE for both configurations, while the gap increases with RF power
% for RF + 4VLP, indicating reduced tightness at high RF SNR.}

Fig.~\ref{SCIM vs RMSE} compares the position and
velocity RMSEs with the corresponding PEM and VEM
derived from the BCRLB and the inverse of the SCIM
for P1 and P2.
For the Monte Carlo evaluation, realizations for P1
are generated according to the assumed Gaussian motion
model, whereas those for P2 are generated by adding
small Gaussian perturbations to the sinusoidal trajectory.
The state expectations in the BIM are approximated
using $100$ Monte Carlo realizations.
The BCRLB is evaluated under the corresponding
true motion model to provide an estimator-independent
performance benchmark.
In contrast, the SCIM retains the Gaussian motion-model
parameters employed by the proposed estimator, {allowing its inverse to serve as an estimator-oriented
performance approximation.}
The BCRLB-based measures remain below the corresponding
RMSEs for both trajectories.
The inverse-SCIM-based measures more closely approximate
the RMSEs, particularly for the model-consistent P1,
in agreement with the MSE relationship established
in Proposition~\ref{prop:SCIM_MSE} under the first-order observation model.
For P2, deviations arise in the presence of motion-model
mismatch, but the inverse-SCIM-based measures still
capture the RMSE trends.
Based on these results, we use the inverse of the SCIM as the primary performance measure in the subsequent analysis.

Figs.~\ref{timeslot} and~\ref{fig:LED_variation}
evaluate the impact of RF measurements on
position and velocity estimation under P1,
using the PEM and VEM computed from the inverse
of the SCIM.
Fig.~\ref{timeslot} compares the PEM, VEM, and PPEM
over time slots for the RF + 4VLP and VLP-only
configurations.
The proposed RF--VLP configuration achieves
lower error measures overall, particularly in VEM.
Its lower PPEM also indicates that the improved
state information contributes to more accurate
position prediction through the state-transition model.
Fig.~\ref{fig:LED_variation} further examines
the impact of RF measurements on the PEM and VEM
as the number of LEDs varies. 
Here, $M$ denotes the number of noncentral LEDs,
so the configuration with a center LED contains
$M+1$ LEDs in total.
For the same $M$, the proposed RF--VLP configuration
achieves lower PEM and VEM than both VLP-only
configurations.
Notably, the proposed RF--VLP configuration with $M=3$
achieves a PEM comparable to those of the VLP-only
configurations with $M=5$ plus a center LED and
$M=7$ noncentral LEDs. 
The greater improvement in VEM across both figures
highlights the contribution of RF Doppler information
to velocity estimation.
% Fig.~\ref{timeslot} shows the PEM, VEM, and PPEM computed from the
% inverse of the SCIM over time slots for P1. The proposed RF + 4VLP
% configuration achieves lower error metrics overall than the VLP-based configurations.
% In particular, its substantially lower VEM demonstrates the benefit of
% RF Doppler information for velocity. The lower PPEM further reflects that the improved state
% information is propagated through the state-transition model,
% thereby improving subsequent position prediction.

{Fig.~\ref{Thorem3_RF variation} shows the PEM and VEM computed from the inverse of the SCIM, together with the corresponding current-slot RF-free baselines and high-SNR limits derived from Theorem~\ref{Theorem: SCIM_RF_gain}. 
For both P1 and P2, the evaluation slot is set to $t=40$.
{The SCIM is recursively computed up to \(t=39\) with the RF transmit
power fixed at \(15\) dBm. At \(t=40\), all other quantities are kept
fixed while the RF transmit power is varied; for the RF-free baseline,
the current-slot RF measurement is omitted.} At low RF power, both error measures decrease as the RF SNR increases, consistent with the first-order RF-induced reduction in the inverse of the SCIM established in Theorem~\ref{Theorem: SCIM_RF_gain}. At high RF power, both measures approach their corresponding asymptotic values derived from  Theorem~\ref{Theorem: SCIM_RF_gain}. 
Fig.~\ref{Thorem4_VLP variation} shows the PEM and VEM
as functions of the VLP transmit power at $t=40$.
The SCIM is recursively computed up to $t=39$ with
the VLP transmit power fixed at $15$ dBm.
{At \(t=40\), we vary only the VLP transmit power while keeping all
other quantities fixed.} {In contrast to the RF case, the PEM approaches
zero as the VLP power increases, whereas the VEM converges to the
nonzero asymptotic value characterized in
Theorem~\ref{theorem:VLP_SCIM_asymptotic}.}}

\section{Conclusion}
In this paper, we investigated joint RF--VLP position
and velocity tracking through Bayesian estimation
and theoretical performance analysis.
We developed a recursive MAP estimator and
characterized the estimation limits and information
gains arising from RF--VLP integration.
Our analysis reveals distinct roles of RF and VLP
measurements in tracking.
While improved VLP measurements enhance position
accuracy, velocity estimation remains limited by
the lack of direct velocity information.
RF Doppler measurements help address this limitation,
providing velocity gains that persist even as the
number of LEDs increases.
The distinct high-SNR behaviors further highlight
the importance of complementary measurement
information in joint position and velocity tracking.
Future work will consider more realistic channel
models, including multipath propagation, and
multiple RF anchors to exploit additional
geometric diversity.
\section*{Appendix A}
\label{CRLB derivate}
Since the considered state-estimation model follows a recursive
Bayesian update, the BIM can be expressed as \cite{recursive_CRLB}
\begin{align}
\mathbf{J}_{t}^{\mathrm{B}}
&=
\mathbf{D}_{t}^{22}
-
\mathbf{D}_{t}^{21}
\left(
\mathbf{J}_{t-1}^{\mathrm{B}}
+
\mathbf{D}_{t}^{11}
\right)^{-1}
\mathbf{D}_{t}^{12},
\label{PCRLB_general_recursion}
\end{align}
where
\begin{subequations}
\label{general_information_submatrices}
\begin{align}
\mathbf{D}_{t}^{11}
&\triangleq
-\mathbb{E}_{\mathbf s_{t-1},\mathbf s_t}\!\left[
\nabla_{\mathbf{s}_{t-1}}
\nabla_{\mathbf{s}_{t-1}}^{T}
\log p(\mathbf{s}_{t}|\mathbf{s}_{t-1})
\right],
\\
\mathbf{D}_{t}^{12}
&\triangleq
-\mathbb{E}_{\mathbf s_{t-1},\mathbf s_t}\!\left[
\nabla_{\mathbf{s}_{t-1}}
\nabla_{\mathbf{s}_{t}}^{T}
\log p(\mathbf{s}_{t}|\mathbf{s}_{t-1})
\right],
\\
\mathbf{D}_{t}^{21}
&\triangleq
-\mathbb{E}_{\mathbf s_{t-1},\mathbf s_t}\!\left[
\nabla_{\mathbf{s}_{t}}
\nabla_{\mathbf{s}_{t-1}}^{T}
\log p(\mathbf{s}_{t}|\mathbf{s}_{t-1})
\right],
\\
\mathbf{D}_{t}^{22}
&\triangleq
-\mathbb{E}_{\mathbf s_{t-1},\mathbf s_t}\!\left[
\nabla_{\mathbf{s}_{t}}
\nabla_{\mathbf{s}_{t}}^{T}
\log p(\mathbf{s}_{t}|\mathbf{s}_{t-1})
\right]
+\mathbf J_t^{\mathrm{meas}}.
\end{align}
\end{subequations}
The expectations are taken over the corresponding state and
observation distributions, where \(\mathbf J_t^{\mathrm{meas}}\) denotes the
equivalent measurement information matrix after eliminating the unknown
RF channel gain. The matrices
\(\mathbf D_{t}^{11}\), \(\mathbf D_{t}^{12}\), and
\(\mathbf D_{t}^{21}\) characterize the state-transition
information, while \(\mathbf D_{t}^{22}\) additionally
includes the current measurement information. For the considered linear Gaussian state-transition model,
\(
p(\mathbf s_t|\mathbf s_{t-1})
=
\mathcal N(
\mathbf s_t;
\mathbf F\mathbf s_{t-1},
\mathbf Q
),
\)
substituting the transition density into
\eqref{general_information_submatrices} yields
\(
\mathbf D_{t}^{11}
=
\mathbf F^T\mathbf Q^{-1}\mathbf F
\),
\(
\mathbf D_{t}^{12}
=
-\mathbf F^T\mathbf Q^{-1}
\),
\(
\mathbf D_{t}^{21}
=
-\mathbf Q^{-1}\mathbf F
\),
and
\(
\mathbf D_{t}^{22}
=
\mathbf Q^{-1}
+
\mathbf J_t^{\mathrm{meas}}.
\) Since the RF and VLP observations are conditionally
independent given \(\mathbf s_t\) and \(\alpha_t\), their joint
likelihood satisfies
\(
p(\mathbf y_t|\mathbf s_t,\alpha_t)
=
p(\mathbf y_t^{\mathrm{RF}}|\mathbf s_t,\alpha_t)
p(\mathbf y_t^{\mathrm{VLP}}|\mathbf s_t).
\)
Accordingly, \(\mathbf J_t^{\mathrm{meas}}\) can be decomposed as
\(
\mathbf J_t^{\mathrm{meas}}
=
\overline{\mathbf J}_t^{\mathrm{RF}}
+
\overline{\mathbf J}_t^{\mathrm{VLP}}.
\)
Consequently,
$\mathbf D_{t}^{22}
=
\mathbf Q^{-1}
+
\mathbf{\bar{J}}_t^{\mathrm{RF}}
+
\mathbf{\bar{J}}_t^{\mathrm{VLP}}.$
Substituting these results into
\eqref{PCRLB_general_recursion} and applying the matrix inversion
lemma \cite{matrix_inversion_Lemma}, the BIM can be expressed as
given in \eqref{Bayesian_information_decomposition}.

Next, we derive the RF information matrix
\(\bar{\mathbf J}_t^{\mathrm{RF}}\)
from the Fisher information matrix (FIM), defined as the
negative expected Hessian of the RF log-likelihood
\cite{estimation_theory}.
Since the RF observation depends on the UE state through delay
and Doppler, we first construct the joint FIM for these parameters
and the unknown complex gain.
We then transform it into the UE-state domain, average its blocks
over the marginal state distribution, and eliminate the unknown gain.
Let
\(\boldsymbol{\psi}_t=[\tau_t,f_{\mathrm D,t}]^T\),
\(\boldsymbol{\alpha}_t
=[\operatorname{Re}\{\alpha_t\},
\operatorname{Im}\{\alpha_t\}]^T\), and
\(\boldsymbol{\theta}_t
=[\boldsymbol{\psi}_t^T,\boldsymbol{\alpha}_t^T]^T\)
denote the delay--Doppler parameter vector, the real-valued
complex gain vector, and the joint parameter vector, respectively.
Under the complex Gaussian RF observation model with
\(\mathbf R^{\mathrm{RF}}=(\sigma^{\mathrm{RF}})^2\mathbf I\),
the conditional FIM for \(\boldsymbol{\theta}_t\) is given by
\cite{estimation_theory}
\begin{align}
\left[\mathbf J_{\boldsymbol{\theta},t}^{\mathrm{RF}}\right]_{p,q}
&=
\frac{2}{(\sigma^{\mathrm{RF}})^{2}}
\operatorname{Re}\!\left\{
\frac{\partial(\boldsymbol{\mu}_t^{\mathrm{RF}})^{H}}
{\partial\theta_{p,t}}
\frac{\partial\boldsymbol{\mu}_t^{\mathrm{RF}}}
{\partial\theta_{q,t}}
\right\},
\label{Gaussian_RF_FIM}
\end{align}
where
\(\boldsymbol{\mu}_t^{\mathrm{RF}}
=\alpha_t\mathbf b_t(\boldsymbol{\psi}_t)\).
The resulting FIM is partitioned as
\begin{align}
\mathbf J_{\boldsymbol{\theta},t}^{\mathrm{RF}}
=
\begin{bmatrix}
\mathbf J_{\boldsymbol{\psi}\boldsymbol{\psi},t}
&
\mathbf J_{\boldsymbol{\psi}\boldsymbol{\alpha},t}
\\
\mathbf J_{\boldsymbol{\alpha}\boldsymbol{\psi},t}
&
\mathbf J_{\boldsymbol{\alpha}\boldsymbol{\alpha},t}
\end{bmatrix},
\label{RF_joint_FIM}
\end{align}
where
\(\mathbf J_{\boldsymbol{\psi}\boldsymbol{\psi},t}
=
\frac{2|\alpha_t|^2}{(\sigma^{\mathrm{RF}})^2}
\operatorname{Re}\{
\mathbf D_{\boldsymbol{\psi},t}^H
\mathbf D_{\boldsymbol{\psi},t}
\}\),
\(\mathbf J_{\boldsymbol{\psi}\boldsymbol{\alpha},t}
=
\frac{2}{(\sigma^{\mathrm{RF}})^2}
\operatorname{Re}\{
\alpha_t^*
\mathbf D_{\boldsymbol{\psi},t}^H
[\mathbf b_t,j\mathbf b_t]
\}\),
\(\mathbf J_{\boldsymbol{\alpha}\boldsymbol{\psi},t}
=
\mathbf J_{\boldsymbol{\psi}\boldsymbol{\alpha},t}^T\), and
\(\mathbf J_{\boldsymbol{\alpha}\boldsymbol{\alpha},t}
=
\frac{2E_b}{(\sigma^{\mathrm{RF}})^2}\mathbf I_2\).
Here,
\(\mathbf D_{\boldsymbol{\psi},t}
\triangleq[\mathbf d_{\tau,t},\mathbf d_{f,t}]\),
with
\(\mathbf d_{\tau,t}
\triangleq\partial\mathbf b_t/\partial\tau_t\)
and
\(\mathbf d_{f,t}
\triangleq\partial\mathbf b_t/\partial f_{\mathrm D,t}\).
From the element-wise expression of \(\mathbf b_t\) in
\eqref{RF_vector_model}, their \(n\)-th elements are respectively given by
\(
[\mathbf d_{\tau,t}]_n
=
-e^{j2\pi f_{\mathrm D,t}\zeta_n}
\dot b(\zeta_n-\tau_t)
\)
and
\(
[\mathbf d_{f,t}]_n
=
j2\pi\zeta_n[\mathbf b_t]_n
\),
where
\(\dot b(\xi)\triangleq\partial b(\xi)/\partial\xi\).
Using the Jacobian transformation from the delay--Doppler domain to
the UE-state domain, the corresponding FIM blocks are given by
\begin{align}
\mathbf J_{ss,t}^{\mathrm{RF}}(\mathbf s_t)
&=
\mathbf H_{\boldsymbol{\psi},t}
\mathbf J_{\boldsymbol{\psi}\boldsymbol{\psi},t}
\mathbf H_{\boldsymbol{\psi},t}^{T},
&
\mathbf J_{s\alpha,t}^{\mathrm{RF}}(\mathbf s_t)
&=
\mathbf H_{\boldsymbol{\psi},t}
\mathbf J_{\boldsymbol{\psi}\boldsymbol{\alpha},t},
\nonumber\\
\mathbf J_{\alpha s,t}^{\mathrm{RF}}(\mathbf s_t)
&=
\mathbf J_{\boldsymbol{\alpha}\boldsymbol{\psi},t}
\mathbf H_{\boldsymbol{\psi},t}^{T},
&
\mathbf J_{\alpha\alpha,t}^{\mathrm{RF}}(\mathbf s_t)
&=
\mathbf J_{\boldsymbol{\alpha}\boldsymbol{\alpha},t}.
\label{RF_state_joint_FIM_blocks}
\end{align}
Here, the delay--Doppler Jacobian is defined as
\begin{align}
    \label{baseline_information_partition}
\mathbf H_{\boldsymbol{\psi},t}
\triangleq
\left(
\frac{\partial\boldsymbol{\psi}_t}
{\partial\mathbf s_t^{T}}
\right)^{T}
=
\begin{bmatrix}
\mathbf r_{\tau,\mathbf p,t}
&
\mathbf r_{f,\mathbf p,t}
\\
\mathbf 0_2
&
\mathbf r_{f,\mathbf v,t}
\end{bmatrix},
\end{align}
where
\(\mathbf r_{\tau,\mathbf p,t}
=
\mathbf u_t^{\mathrm{RF}}/c_0\),
\(\mathbf r_{f,\mathbf p,t}
=
-[\mathbf I_2-\mathbf u_t^{\mathrm{RF}}
(\mathbf u_t^{\mathrm{RF}})^T]\mathbf v_t/
(\lambda d_t^{\mathrm{RF}})\),
and
\(\mathbf r_{f,\mathbf v,t}
=
-\mathbf u_t^{\mathrm{RF}}/\lambda\). To obtain the Bayesian RF information contribution, the above
state-dependent FIM blocks are averaged over the marginal state
distribution of \(\mathbf s_t\) as
\(\overline{\mathbf J}_{ab,t}^{\mathrm{RF}}
\triangleq
\mathbb E_{\mathbf s_t}
[\mathbf J_{ab,t}^{\mathrm{RF}}(\mathbf s_t)]\),
where \(ab\in\{ss,s\alpha,\alpha s,\alpha\alpha\}\).
Eliminating the unknown complex gain \(\boldsymbol{\alpha}_t\) from
the averaged joint FIM through the Schur complement yields the
equivalent Bayesian RF information matrix for the UE state, given by
\eqref{RF_information_matrix}.

Next, we derive the VLP information contribution
\(\bar{\mathbf J}_t^{\mathrm{VLP}}\).
The VLP observation mean gradients can be obtained from
\eqref{VLP_residual_derivative}.
Moreover, the observations from different LEDs are conditionally
independent given \(\mathbf s_t\), allowing their information
contributions to be summed. Averaging the resulting FIM over the marginal
state distribution yields
\begin{align}
\bar{\mathbf J}_t^{\mathrm{VLP}}
&=
\mathbb E_{\mathbf s_t}\!\left[
\frac{1}{(\sigma^{\mathrm{VLP}})^{2}}
\sum_{m=1}^{M}
\nabla_{\mathbf s_t}\mu_{m,t}^{\mathrm{VLP}}
\left(
\nabla_{\mathbf s_t}\mu_{m,t}^{\mathrm{VLP}}
\right)^{T}
\right]
\nonumber\\
&=
\frac{P^{\mathrm{LED}}}{(\sigma^{\mathrm{VLP}})^{2}}
\sum_{m=1}^{M}
\mathbb E_{\mathbf s_t}\!\left[
\nabla_{\mathbf s_t}h_{m,t}^{\mathrm{VLP}}
\left(
\nabla_{\mathbf s_t}h_{m,t}^{\mathrm{VLP}}
\right)^{T}
\right].
\label{VLP_FIM_general}
\end{align}
%Using the derivatives in \eqref{VLP_derivate}, the channel-gradient vector is expressed as
%\begin{align}
%\nabla_{\mathbf{s}_{t}}h_{m,t}^{\mathrm{VLP}}
%&=
%-\frac{(r+3)h_{m,t}^{\mathrm{VLP}}}{d_{m,t}^{2}}
%\begin{bmatrix}
%\mathbf p_t-\mathbf p^{\mathrm{LED}}_m
%\\
%\mathbf 0_2
%\end{bmatrix}.
%\label{VLP_channel_gradient_vector}
%\end{align}
%Substituting \eqref{VLP_channel_gradient_vector} into \eqref{VLP_FIM_general} yields the Bayesian VLP information contribution in \eqref{VLP_information_matrix}.
Using the derivatives in \eqref{VLP_derivate} and substituting them into \eqref{VLP_FIM_general} yields the Bayesian VLP information contribution in \eqref{VLP_information_matrix}.
\section*{Appendix B}
%As noted in the proof of Lemma~\ref{lemma:conditional_information}, the VLP and prediction terms follow directly from Theorem~\ref{theorem:BIM_terms} under state conditioni44ng. Therefore, it remains to derive the state-conditioned RF information matrix \(\mathbf J_t^{\mathrm{RF,c}}(\mathbf s_t)\).
Following the RF derivation in Appendix~A, we consider the
joint RF information matrix in \eqref{RF_joint_FIM} without
averaging over the UE state. Eliminating the unknown RF channel
gain through the Schur complement yields the equivalent
delay--Doppler information matrix for
\(\boldsymbol{\psi}_t\) as
\begin{align}
\mathbf{J}_{\boldsymbol{\psi},t}^{\mathrm{eq}}
&=
\mathbf{J}_{\boldsymbol{\psi}\boldsymbol{\psi},t}
-
\mathbf{J}_{\boldsymbol{\psi}\boldsymbol{\alpha},t}
\mathbf{J}_{\boldsymbol{\alpha}\boldsymbol{\alpha},t}^{-1}
\mathbf{J}_{\boldsymbol{\alpha}\boldsymbol{\psi},t}
\nonumber\\
&\overset{(a)}{=}
\frac{2|\alpha_t|^{2}}
{(\sigma^{\mathrm{RF}})^2}
\operatorname{Re}\!\left\{
\mathbf{D}_{\boldsymbol{\psi},t}^{H}
\mathbf{\Pi}_{\mathbf b_t}^{\perp}
\mathbf{D}_{\boldsymbol{\psi},t}
\right\}
\nonumber\\
&\overset{(b)}{=}
\frac{2|\alpha_t|^{2}}
{(\sigma^{\mathrm{RF}})^2}
\operatorname{Re}\!\left\{
\mathbf{D}_{\boldsymbol{\psi},t}^{H}
\mathbf{D}_{\boldsymbol{\psi},t}
-
\frac{
\mathbf{D}_{\boldsymbol{\psi},t}^{H}
\mathbf b_t\mathbf b_t^{H}
\mathbf{D}_{\boldsymbol{\psi},t}
}{E_b}
\right\},
\label{RF_EFIM_psi}
\end{align}
where
\(\mathbf{\Pi}_{\mathbf b_t}^{\perp}
\triangleq\mathbf I-\mathbf b_t\mathbf b_t^{H}/E_b\).
Here, $(a)$ follows from the projection-form equivalent Fisher information matrix (EFIM) in
\cite{CRLB_projection}, while $(b)$ follows directly from the definition
of \(\mathbf{\Pi}_{\mathbf{b}_t}^{\perp}\). Define the discrete-sample effective bandwidth, effective time duration,
and normalized delay--Doppler coupling coefficient \cite{effective_BW} as
$\beta^{2}
=  
\left[
\frac{
\sum_{n=1}^{N^{\mathrm{RF}}}
|\dot b(\zeta_n-\tau_t)|^{2}
}{4\pi^{2}E_b}
-
\frac{
\left|
\sum_{n=1}^{N^{\mathrm{RF}}}
\dot b^{*}(\zeta_n-\tau_t)b(\zeta_n-\tau_t)
\right|^{2}
}{4\pi^{2}E_b^{2}}
\right],
T_{\mathrm{eff}}^{2}
\triangleq
\frac{
\sum_{n=1}^{N^{\mathrm{RF}}}
\zeta_n^{2}|b_{t,n}|^{2}
}{E_b}
-
\left(
\frac{
\sum_{n=1}^{N^{\mathrm{RF}}}
\zeta_n|b_{t,n}|^{2}
}{E_b}
\right)^{2},
$ and $\rho_{\beta T}
\triangleq
\frac{
\operatorname{Re}
\left\{
\mathbf d_{\tau,t}^{H}
\mathbf{\Pi}_{\mathbf{b}_t}^{\perp}
\mathbf d_{f,t}
\right\}
}{
4\pi^{2}E_b\beta T_{\mathrm{eff}}
}.
$ Using these, the EFIM with respect to
\(\boldsymbol{\psi}_t\) can be  expressed as
$
\mathbf{J}_{\boldsymbol{\psi},t}^{\mathrm{eq}}
=
8\pi^2{\Gamma}_{t}^{\mathrm{RF}}
\begin{bmatrix}
\beta^{2}
&
\rho_{\beta T}\beta T_{\mathrm{eff}}
\\
\rho_{\beta T}\beta T_{\mathrm{eff}}
&
T_{\mathrm{eff}}^{2}
\end{bmatrix}.
$
Finally, \(\mathbf{J}_{t}^{\mathrm{RF,c}}(\mathbf s_t)\) is obtained
through the Jacobian transformation as given in~\eqref{conditional_RF_information}.

\section*{Appendix C}
From the SCIM decomposition, its block matrices are given by
\(
\mathbf A_t
=
\mathbf A_{0,t}
+
\widetilde{\Gamma}_{t}^{\mathrm{RF}}
\beta^{2}(1-\rho_{\beta T}^{2})
\mathbf r_{\tau,\mathbf p,t}
\mathbf r_{\tau,\mathbf p,t}^{T}
+
\kappa_t\mathbf g_t\mathbf g_t^{T}
\),
\(
\mathbf C_t
=
\mathbf C_{0,t}
+
\kappa_t\mathbf g_t
\mathbf r_{f,\mathbf v,t}^{T}
\),
and
\(
\mathbf D_t
=
\mathbf D_{0,t}
+
\kappa_t
\mathbf r_{f,\mathbf v,t}
\mathbf r_{f,\mathbf v,t}^{T}
\),
where
\(
\kappa_t
\triangleq
\widetilde{\Gamma}_{t}^{\mathrm{RF}}T_{\mathrm{eff}}^{2}
\).
Applying the Sherman--Morrison identity to \(\mathbf D_t^{-1}\) gives
\(
\mathbf D_t^{-1}
=
\mathbf D_{0,t}^{-1}
-
\frac{
\kappa_t
\mathbf D_{0,t}^{-1}
\mathbf r_{f,\mathbf v,t}
\mathbf r_{f,\mathbf v,t}^{T}
\mathbf D_{0,t}^{-1}
}{
1+\kappa_t a_t
}
\),
where
\(
a_t
\triangleq
\mathbf r_{f,\mathbf v,t}^{T}
\mathbf D_{0,t}^{-1}
\mathbf r_{f,\mathbf v,t}
\).
Substituting these expressions into
\(\mathbf E_{\mathbf p,t}
=
\mathbf A_t-\mathbf C_t\mathbf D_t^{-1}\mathbf C_t^{T}\)
and subtracting \(\mathbf E_{\mathbf p,0,t}\) yields the RF-induced
position-SCIM gain in \eqref{RF_information_gain}.

We now analyze the low- and high-SNR behaviors of
\(\Delta\mathbf J_{\mathbf p,t}^{\mathrm{RF}}\).
In the low-SNR regime, i.e., as
\(\widetilde{\Gamma}_{t}^{\mathrm{RF}}\rightarrow0\),
the coefficient of the Doppler-induced gain satisfies
$\frac{
\widetilde{\Gamma}_{t}^{\mathrm{RF}}T_{\mathrm{eff}}^{2}
}{
1+\widetilde{\Gamma}_{t}^{\mathrm{RF}}
T_{\mathrm{eff}}^{2}a_t
}
=
\widetilde{\Gamma}_{t}^{\mathrm{RF}}T_{\mathrm{eff}}^{2}
-
(\widetilde{\Gamma}_{t}^{\mathrm{RF}})^{2}
T_{\mathrm{eff}}^{4}a_t
+
\mathcal O((\widetilde{\Gamma}_{t}^{\mathrm{RF}})^{3})$. 
Since the delay-induced gain is linear in
\(\widetilde{\Gamma}_{t}^{\mathrm{RF}}\), substituting this  
expansion into \eqref{RF_information_gain} yields
\eqref{RF_gain_low_SNR}.

In the high-SNR regime, i.e., as
\(\widetilde{\Gamma}_{t}^{\mathrm{RF}}\rightarrow\infty\),
the coefficient of the Doppler-induced gain satisfies
$\frac{
\widetilde{\Gamma}_{t}^{\mathrm{RF}}T_{\mathrm{eff}}^{2}
}{
1+\widetilde{\Gamma}_{t}^{\mathrm{RF}}
T_{\mathrm{eff}}^{2}a_t
}=
\frac{1}{a_t}
+
\mathcal O((\widetilde{\Gamma}_{t}^{\mathrm{RF}})^{-1})$. 
Since the delay-induced gain is linear in
\(\widetilde{\Gamma}_{t}^{\mathrm{RF}}\), substituting the above
expansion into \eqref{RF_information_gain} yields
\eqref{RF_gain_high_SNR}.
\section*{Appendix D}
Using the SCIM blocks given in Appendix C and applying the Sherman–Morrison identity to $\mathbf{A}_t=\mathbf{A}_{\tau,t}+\kappa_t \mathbf{g}_t \mathbf{g}_t^T$, the equivalent velocity-information matrix can be expressed as
\begin{align}
\mathbf E_{\mathbf v,t}
&=
\mathbf D_{0,t}
-
\mathbf C_{0,t}^{T}
\mathbf A_{\tau,t}^{-1}
(\widetilde{\Gamma}_{t}^{\mathrm{RF}})
\mathbf C_{0,t}
\nonumber\\
&\quad+
\frac{
\widetilde{\Gamma}_{t}^{\mathrm{RF}}T_{\mathrm{eff}}^{2}
\mathbf z_{v,t}(\widetilde{\Gamma}_{t}^{\mathrm{RF}})
\mathbf z_{v,t}^{T}(\widetilde{\Gamma}_{t}^{\mathrm{RF}})
}{
1+\widetilde{\Gamma}_{t}^{\mathrm{RF}}T_{\mathrm{eff}}^{2}
b_{v,t}(\widetilde{\Gamma}_{t}^{\mathrm{RF}})
}.
\label{RF_velocity_EFIM}
\end{align}
Using the definition of \(\mathbf A_{\tau,t}\) in
Theorem~\ref{theorem:RF_velocity_gain} and applying the
Sherman--Morrison identity, we obtain
\(
\mathbf C_{0,t}^{T}
\mathbf A_{\tau,t}^{-1}
(\widetilde{\Gamma}_{t}^{\mathrm{RF}})
\mathbf C_{0,t}
=
\mathbf C_{0,t}^{T}
\mathbf A_{0,t}^{-1}
\mathbf C_{0,t}
-
\frac{
\widetilde{\Gamma}_{t}^{\mathrm{RF}}\delta_{\tau}
}{
1+\widetilde{\Gamma}_{t}^{\mathrm{RF}}\delta_{\tau}c_t
}
\mathbf w_t\mathbf w_t^{T}
\).
Substituting this result into \eqref{RF_velocity_EFIM}
and subtracting the RF-free equivalent velocity-information matrix
\(\mathbf E_{\mathbf v,0,t}\) yields
\eqref{RF_velocity_information_gain}.

We now analyze the low- and high-SNR behaviors of
\eqref{RF_velocity_information_gain}. In the low-SNR regime, i.e., as
\(\widetilde{\Gamma}_{t}^{\mathrm{RF}}\rightarrow0\), the
delay-induced term in \eqref{RF_velocity_information_gain} satisfies
\begin{align}
\frac{
\widetilde{\Gamma}_{t}^{\mathrm{RF}}\delta_{\tau}
}{
1+\widetilde{\Gamma}_{t}^{\mathrm{RF}}\delta_{\tau}c_t
}
\mathbf w_t\mathbf w_t^{T}
&=
\widetilde{\Gamma}_{t}^{\mathrm{RF}}\delta_{\tau}
\mathbf w_t\mathbf w_t^{T}
+
\mathcal O((\widetilde{\Gamma}_{t}^{\mathrm{RF}})^2).
\label{RF_velocity_low_SNR_delay}
\end{align} 
Next, for the Doppler-induced term, as
\(\widetilde{\Gamma}_{t}^{\mathrm{RF}}\rightarrow0\),
\(\mathbf z_{v,t}(\widetilde{\Gamma}_{t}^{\mathrm{RF}})
=
\mathbf z_{v,t}^{0}
+
\widetilde{\Gamma}_{t}^{\mathrm{RF}}
\delta_{\tau}\chi_t\mathbf w_t
+
\mathcal O((\widetilde{\Gamma}_{t}^{\mathrm{RF}})^2)\)
and
\(b_{v,t}(\widetilde{\Gamma}_{t}^{\mathrm{RF}})
=
b_{v,t}^{0}
-
\widetilde{\Gamma}_{t}^{\mathrm{RF}}
\delta_{\tau}\chi_t^{2}
+
\mathcal O((\widetilde{\Gamma}_{t}^{\mathrm{RF}})^2)\),
where
\(\mathbf z_{v,t}^{0}= \mathbf r_{f,v,t}
-\mathbf C_{0,t}^{T}\mathbf A_{0,t}^{-1}\mathbf g_t\),
\(b_{v,t}^{0}=
\mathbf g_t^{T}\mathbf A_{0,t}^{-1}\mathbf g_t\), and
\(\chi_t=
\mathbf r_{\tau,p,t}^{T}\mathbf A_{0,t}^{-1}\mathbf g_t\).
Using these expansions, the coefficient of the
Doppler-induced term satisfies
\begin{align}
\frac{
\widetilde{\Gamma}_{t}^{\mathrm{RF}}T_{\mathrm{eff}}^{2}
}{
1\!+\!\widetilde{\Gamma}_{t}^{\mathrm{RF}}T_{\mathrm{eff}}^{2}
b_{v,t}(\widetilde{\Gamma}_{t}^{\mathrm{RF}})
}
&\!=\!
\widetilde{\Gamma}_{t}^{\mathrm{RF}}T_{\mathrm{eff}}^{2}
\!-\!
(\widetilde{\Gamma}_{t}^{\mathrm{RF}})^2
T_{\mathrm{eff}}^{4}b_{v,t}^{0}
\!+\!
\mathcal O((\widetilde{\Gamma}_{t}^{\mathrm{RF}})^3)
\nonumber\\
&\!=\!
\widetilde{\Gamma}_{t}^{\mathrm{RF}}T_{\mathrm{eff}}^{2}
+
\mathcal O((\widetilde{\Gamma}_{t}^{\mathrm{RF}})^2).
\label{RF_velocity_low_SNR_coefficient}
\end{align}
Moreover,
\(\mathbf z_{v,t}(\widetilde{\Gamma}_{t}^{\mathrm{RF}})
\mathbf z_{v,t}^{T}(\widetilde{\Gamma}_{t}^{\mathrm{RF}})
=
\mathbf z_{v,t}^{0}(\mathbf z_{v,t}^{0})^{T}
+
\mathcal O(\widetilde{\Gamma}_{t}^{\mathrm{RF}})\).
Combining this relation with
\eqref{RF_velocity_low_SNR_delay} and
\eqref{RF_velocity_low_SNR_coefficient} yields
\eqref{RF_velocity_gain_low_SNR}.

In the high-SNR regime, i.e., as
\(\widetilde{\Gamma}_{t}^{\mathrm{RF}}\rightarrow\infty\),
$
b_{v,t}(\widetilde{\Gamma}_{t}^{\mathrm{RF}})
\rightarrow
b_{v,t}^{\infty}
\triangleq
b_{v,t}^{0}-\frac{\chi_t^{2}}{c_t},
$ and $
\mathbf z_{v,t}(\widetilde{\Gamma}_{t}^{\mathrm{RF}})
\rightarrow
\mathbf z_{v,t}^{\infty}
\triangleq
\mathbf z_{v,t}^{0}
+\frac{\chi_t}{c_t}\mathbf w_t.
$
Moreover,
\(\frac{\widetilde{\Gamma}_{t}^{\mathrm{RF}}\delta_{\tau}}
{1+\widetilde{\Gamma}_{t}^{\mathrm{RF}}\delta_{\tau}c_t}
\rightarrow\frac{1}{c_t},
\qquad
\frac{\widetilde{\Gamma}_{t}^{\mathrm{RF}} T_{\mathrm{eff}}^{2}}
{1+\widetilde{\Gamma}_{t}^{\mathrm{RF}} T_{\mathrm{eff}}^{2}b_{v,t}(\widetilde{\Gamma}_{t}^{\mathrm{RF}})}
\rightarrow\frac{1}{b_{v,t}^{\infty}}.
\) Consequently, the RF-induced velocity-information gain approaches the
high-SNR limit in \eqref{RF_velocity_gain_high_SNR}.

\section*{Appendix E}
Recall that the RF-free baseline SCIM is defined as
\(
\boldsymbol{\Lambda}_{0,t}
=
\mathbf J_t^{\mathrm{VLP,c}}
+
\mathbf J_t^{\mathrm{pred,c}}.
\)
Then, from \eqref{conditional_information_decomposition}, the SCIM is given by
\(
\mathbf J_t^{\mathrm c}
=
\boldsymbol{\Lambda}_{0,t}
+
\widetilde{\Gamma}_{t}^{\mathrm{RF}}
\mathbf H_{\boldsymbol{\psi},t}
\mathbf K_{\boldsymbol{\psi}}
\mathbf H_{\boldsymbol{\psi},t}^{T}.
\)
It follows that
\begin{align}
(\mathbf J_t^{\mathrm c})^{-1}
&=
\left[
\boldsymbol{\Lambda}_{0,t}
+
\widetilde{\Gamma}_{t}^{\mathrm{RF}}
\mathbf H_{\boldsymbol{\psi},t}
\mathbf K_{\boldsymbol{\psi}}
\mathbf H_{\boldsymbol{\psi},t}^{T}
\right]^{-1}.
\label{RF_SCIM_proof}
\end{align}
Applying the Woodbury matrix identity \cite{golub2013matrix} to \eqref{RF_SCIM_proof} yields \eqref{RF_CRLB_gain}.
We next characterize the asymptotic behavior of
\(\Delta\mathbf P_t^{\mathrm{RF}}\) with respect to the RF SNR.

In the low-SNR regime, i.e.,
\(\widetilde{\Gamma}_{t}^{\mathrm{RF}}\rightarrow0\), the inverse term in
\eqref{RF_CRLB_gain} can be written as
$\left[
(\widetilde{\Gamma}_{t}^{\mathrm{RF}})^{-1}
\mathbf K_{\boldsymbol{\psi}}^{-1}
+
\mathbf H_{\boldsymbol{\psi},t}^{T}
\boldsymbol{\Lambda}_{0,t}^{-1}
\mathbf H_{\boldsymbol{\psi},t}
\right]^{-1}\!\!\!=
\!
\widetilde{\Gamma}_{t}^{\mathrm{RF}}
\mathbf K_{\boldsymbol{\psi}}
\!+\!
\mathcal O\!\left(
(\widetilde{\Gamma}_{t}^{\mathrm{RF}})^2
\right)$. 
Substituting this expansion into \eqref{RF_CRLB_gain} yields
\eqref{RF_CRLB_low_SNR}.
In the high-SNR regime,  the term
\((\widetilde{\Gamma}_{t}^{\mathrm{RF}})^{-1}
\mathbf K_{\boldsymbol{\psi}}^{-1}\) in \eqref{RF_CRLB_gain}
vanishes. Therefore, the inverse term converges to
\(
[
\mathbf H_{\boldsymbol{\psi},t}^{T}
\boldsymbol{\Lambda}_{0,t}^{-1}
\mathbf H_{\boldsymbol{\psi},t}
]^{-1}
\),
which directly yields \eqref{RF_CRLB_high_SNR}.
\section*{Appendix F}
From the definition of $\mathbf{\Lambda}_t^{\text{RP}}$, the SCIM can be expressed as
\begin{align}
\mathbf J_t^{\mathrm c}
&=
\begin{bmatrix}
\mathbf A_t^{\mathrm{RP}}
+
\Gamma_t^{\mathrm{VLP}}
\mathbf G_t^{\mathrm{VLP}}
&
\mathbf C_t^{\mathrm{RP}}
\\
(\mathbf C_t^{\mathrm{RP}})^T
&
\mathbf D_t^{\mathrm{RP}}
\end{bmatrix}.
\end{align}
Applying the block matrix inversion formula yields
\begin{align}
(\mathbf J_t^{\mathrm c})^{-1}
&=
\begin{bmatrix}
\boldsymbol{\Xi}_t^{-1}
&
-\boldsymbol{\Xi}_t^{-1}
\mathbf C_t^{\mathrm{RP}}
(\mathbf D_t^{\mathrm{RP}})^{-1}
\\
-(\mathbf D_t^{\mathrm{RP}})^{-1}
(\mathbf C_t^{\mathrm{RP}})^T
\boldsymbol{\Xi}_t^{-1}
&
\boldsymbol{\Omega}_t
\end{bmatrix},
\label{VLP_CRLB_exact}
\end{align}
where
\(
\boldsymbol{\Xi}_t
\triangleq
\Gamma_t^{\mathrm{VLP}}\mathbf G_t^{\mathrm{VLP}}
+\mathbf B_t^{\mathrm{RP}}
\), \(
\mathbf B_t^{\mathrm{RP}}
\triangleq
\mathbf A_t^{\mathrm{RP}}
-
\mathbf C_t^{\mathrm{RP}}
(\mathbf D_t^{\mathrm{RP}})^{-1}
(\mathbf C_t^{\mathrm{RP}})^T
\)
and
\(
\boldsymbol{\Omega}_t
\triangleq
(\mathbf D_t^{\mathrm{RP}})^{-1}
+
(\mathbf D_t^{\mathrm{RP}})^{-1}
(\mathbf C_t^{\mathrm{RP}})^T
\boldsymbol{\Xi}_t^{-1}
\mathbf C_t^{\mathrm{RP}}
(\mathbf D_t^{\mathrm{RP}})^{-1}
\).

For the low-SNR regime, as
\(\Gamma_t^{\mathrm{VLP}}\rightarrow0\), the inverse of
\(\boldsymbol{\Xi}_t\) admits the expansion
\(
\boldsymbol{\Xi}_t^{-1}
=
(\mathbf B_t^{\mathrm{RP}})^{-1}
-
\Gamma_t^{\mathrm{VLP}}
(\mathbf B_t^{\mathrm{RP}})^{-1}
\mathbf G_t^{\mathrm{VLP}}
(\mathbf B_t^{\mathrm{RP}})^{-1}
+
\mathcal O((\Gamma_t^{\mathrm{VLP}})^2)
\).
Applying the block matrix inversion formula to
\(\boldsymbol{\Lambda}_t^{\mathrm{RP}}\) gives
$
(\boldsymbol{\Lambda}_t^{\mathrm{RP}})^{-1}=
\begin{bmatrix}
(\mathbf B_t^{\mathrm{RP}})^{-1}
&
\mathbf P_{12,t}^{\mathrm{RP}}
\\
(\mathbf P_{12,t}^{\mathrm{RP}})^T
&
\mathbf \Omega_t^{\mathrm{RP}}
\end{bmatrix},
$
where
\(
\mathbf P_{12,t}^{\mathrm{RP}}
\triangleq
-(\mathbf B_t^{\mathrm{RP}})^{-1}
\mathbf C_t^{\mathrm{RP}}
(\mathbf D_t^{\mathrm{RP}})^{-1}
\)
and
\(
\mathbf \Omega_t^{\mathrm{RP}}
\triangleq
(\mathbf D_t^{\mathrm{RP}})^{-1}
+
(\mathbf D_t^{\mathrm{RP}})^{-1}
(\mathbf C_t^{\mathrm{RP}})^T
(\mathbf B_t^{\mathrm{RP}})^{-1}
\mathbf C_t^{\mathrm{RP}}
(\mathbf D_t^{\mathrm{RP}})^{-1}
\).
Substituting the expansion of \(\boldsymbol{\Xi}_t^{-1}\) into
\eqref{VLP_CRLB_exact} and retaining terms up to first order yields
\eqref{VLP_CRLB_low_SNR}.

For the high-SNR regime, the matrix inverse expansion gives
\(
\boldsymbol{\Xi}_t^{-1}
=
\frac{1}{\Gamma_t^{\mathrm{VLP}}}
(\mathbf G_t^{\mathrm{VLP}})^{-1}
-
\frac{1}{(\Gamma_t^{\mathrm{VLP}})^2}
(\mathbf G_t^{\mathrm{VLP}})^{-1}
\mathbf B_t^{\mathrm{RP}}
(\mathbf G_t^{\mathrm{VLP}})^{-1}
+
\mathcal O\!\left(
(\Gamma_t^{\mathrm{VLP}})^{-3}
\right)
\).
Substituting this expansion into \eqref{VLP_CRLB_exact} and using the
definitions of \(\mathbf Z_t\) and \(\mathbf W_t\) in
Theorem~\ref{theorem:VLP_SCIM_asymptotic} yields
\eqref{VLP_CRLB_high_SNR}. Consequently, the limits of the position and velocity blocks in \eqref{VLP_CRLB_limits} follow directly, which
completes the proof.

% Can use something like this to put references on a page
% by themselves when using endfloat and the captionsoff option.
\ifCLASSOPTIONcaptionsoff
  \newpage
\fi

% trigger a \newpage just before the given reference
% number - used to balance the columns on the last page
% adjust value as needed - may need to be readjusted if
% the document is modified later
%\IEEEtriggeratref{8}
% The "triggered" command can be changed if desired:
%\IEEEtriggercmd{\enlargethispage{-5in}}

% references section
%\bibliographystyle{unsrt}
 \vspace{-0.2cm}
% can use a bibliography generated by BibTeX as a .bbl file
% BibTeX documentation can be easily obtained at:
% http://mirror.ctan.org/biblio/bibtex/contrib/doc/
% The IEEEtran BibTeX style support page is at:
% http://www.michaelshell.org/tex/ieeetran/bibtex/
\bibliographystyle{IEEEtran}
\bibliography{bare_jrnl}
% argument is your BibTeX string definitions and bibliography database(s)
%\bibliography{IEEEabrv,../bib/paper}
%
% <OR> manually copy in the resultant .bbl file
% set second argument of \begin to the number of references
% (used to reserve space for the reference number labels box)

% biography section
% 
% If you have an EPS/PDF photo (graphicx package needed) extra braces are
% needed around the contents of the optional argument to biography to prevent
% the LaTeX parser from getting confused when it sees the complicated
% \includegraphics command within an optional argument. (You could create
% your own custom macro containing the \includegraphics command to make things
% simpler here.)
%\begin{IEEEbiography}[{\includegraphics[width=1in,height=1.25in,clip,keepaspectratio]{mshell}}]{Michael Shell}
% or if you just want to reserve a space for a photo:

% You can push biographies down or up by placing
% a \vfill before or after them. The appropriate
% use of \vfill depends on what kind of text is
% on the last page and whether or not the columns
% are being equalized.

%\vfill

% Can be used to pull up biographies so that the bottom of the last one
% is flush with the other column.
%\enlargethispage{-5in}

% that's all folks
\end{document}